\documentclass[preprint,11pt]{aastex}
\usepackage{emulateapj5}

\newcommand{\ee}[1]{\mbox{${} \times 10^{#1}$}}
\newcommand{\eten}[1]{\mbox{$10^{#1}$}}

\def\HII {\hbox{H\,{\scriptsize II}}}
\def\CP {\hbox{C$^+$}}
\def\CZ {\hbox{C$^0$}}

\def\HCOP {\hbox{HCO$^+$}}
\def\CO {\hbox{$^{12}$CO}}
\def\CCO {\hbox{$^{13}$CO}}
\def\COO {\hbox{C$^{18}$O}}
\newcommand{\hh}{\mbox{{\rm H}$_2$}}

\newcommand\CII {\mbox{[C\,{\scriptsize II}]}}
\newcommand\OI {\mbox{[O\,{\scriptsize I}]}}
\newcommand\CI {\mbox{[C\,{\scriptsize I}]}}
\def\COTWO {\CO\,$J\!=\!2\!\rightarrow\!1$}
\def\CCOONE {\CCO\,$J\!=\!1\!\rightarrow\!0$}
\def\CCOTWO {\CCO\,$J\!=\!2\!\rightarrow\!1$}
\def\COOTWO {\COO\,$J\!=\!2\!\rightarrow\!1$}
\def\CCOTHR {\CCO\,$J\!=\!3\!\rightarrow\!2$}
\def\HCOPTHR {\HCOP\,$J\!=\!3\!\rightarrow\!2$}

\def\h {$^{\rm h}$}
\def\m {$^{\rm m}$}
\def\s {$^{\rm s}$}
\newcommand{\degree}{\mbox{$^\circ$}}

\def\cma {\hbox{cm$^{-2}$}}
\def\cmv {\hbox{cm$^{-3}$}}
\def\kms {\hbox{km\,s$^{-1}$}}

\def\AV {\hbox{$A_V$}}

\newcommand{\fir}{\mbox{far-infrared}}
\newcommand{\submm}{\mbox{submillimeter}}
\newcommand{\fuv}{\mbox{far-ultraviolet}}
\newcommand{\iso}{\mbox{\it{ISO}}}
\newcommand{\go}{\mbox{$G_0$}}

\newcommand{\icgs}{\mbox{ergs s$^{-1}$ cm$^{-2}$ sr$^{-1}$}}
\newcommand{\jj}[2]{\mbox{$J = #1\rightarrow#2$}}
\newcommand{\xe}{\mbox{$x(e)$}}
\newcommand{\tad}{\mbox{$t_{AD}$}}
\newcommand{\tk}{\mbox{$T_K$}}

\slugcomment{\footnotesize {\LaTeX}ed at \number\time\ min., \today}
\shorttitle{Peripheral Region of L1204/S140}
\shortauthors{Li et al.}

\begin{document}

\title{Photon Dominated Regions in Low UV Fields:
A Study of The Peripheral Region of L1204/S140}

\author{Wenbin Li}
\affil{Department of Astronomy, The University of Texas at Austin\\
       Austin, TX 78712\\
       and \\
       Enterprise Services, Sun Microsystems of Canada, Inc.\\
       15 Allstate Parkway, Suite \#300, \\
       Markham, ONT L3R 5B4, Canada}
\email{wenbin.li@canada.sun.com}
\author{Neal J. Evans II}
\affil{Department of Astronomy, The University of Texas at Austin\\
       Austin, TX 78712, and Sterrewacht Leiden, P.O.Box 9513, 2300 RA Leiden,
       the Netherlands}
\email{nje@astro.as.utexas.edu}
\author{Daniel T. Jaffe}
\affil{Department of Astronomy, The University of Texas at Austin\\
       Austin, TX 78712}
\email{dtj@astro.as.utexas.edu}
\author{Ewine F. van Dishoeck\altaffilmark{1}}
\affil{Sterrewacht Leiden, P.O.Box 9513, 2300 RA Leiden, the Netherlands}
\email{ewine@strw.LeidenUniv.nl}
\author{Wing-Fai Thi}
\affil{Sterrewacht Leiden, P.O.Box 9513, 2300 RA Leiden, the Netherlands}
\email{thi@strw.LeidenUniv.nl}

\altaffiltext{1}{The Beatrice M. Tinsley Centennial Visiting Professor,
the University of Texas at Austin.}


\begin{abstract}

We have carried out an in-depth study of the peripheral
region of the molecular cloud L1204/S140, where the \fuv\ radiation
and the density are relatively low.
Our observations test theories  of photon-dominated regions (PDRs)
in a regime that has been little explored.  Knowledge of such regions will 
also help
to test theories of photoionization-regulated star formation.
\CII\ 158 \micron\ and \OI\ 63 \micron\ lines are detected by ISO
at all 16 positions along a 1-dimensional cut in right ascension.
Emission from \hh\ rotational transitions \jj20\ and \jj31, at
28 and 17 \micron, was also detected at several positions.
The \CII, \OI, and \hh\ intensities along the
cut show much less spatial variation than do the
rotational lines of \CO\ and other CO isotopes.
The average \CII\ and \OI\ intensities and their ratio
are consistent with models of PDRs with low \fuv\ radiation (\go) and density.
The best-fitting model has $\go  \sim 15$ and  density, $n \sim \eten{3}$ 
\cmv.
Standard PDR models underpredict the intensity in the \hh\ rotational
lines by up to an order of magnitude.
This problem has also been seen in bright PDRs and attributed to
factors, such as geometry and gas-grain drift, that should be much
less important in the regime studied here. The fact that we see the same
problem
in our data suggests that more fundamental solutions, such as higher
\hh\ formation rates, are needed.
Also, in this regime of low density and small line width, the \OI\ line is
sensitive to the radiative transfer and geometry.
Using the ionization structure of the models, a
quantitative analysis of timescales for ambipolar diffusion
in the peripheral regions of the S140 cloud
is consistent with a 
theory of photoionization-regulated star formation.
Observations of \CII\ in other galaxies differ both from those of
high \go\ PDRs in our galaxy and from the low \go\ regions we have studied. 
The extragalactic results are not easily reproduced with mixtures of
high and low \go\ regions.

\end{abstract}


\section{INTRODUCTION}

Photon-dominated regions (or photodissociation regions, PDRs)
are regions of the neutral interstellar medium (ISM) where far-ultraviolet 
(FUV)
(6 eV $< h\nu < 13.6$ eV) photons control the heating and chemical processes.
They are the interface between \HII\ regions and cold molecular cores.
The physical and chemical structure of PDRs depends critically on the
FUV intensity and the gas density.
In the peripheral regions of the molecular cloud L1204/S140, the FUV
intensity and the gas density are low, allowing tests of the models
in an important regime.
We are motivated by two primary goals:
understanding PDRs in a regime of parameter space that has not been
extensively studied; and understanding the role of the regions with
relatively low FUV and density in global star formation, both in our
own Galaxy and in other galaxies.

A great deal of observational and theoretical
effort has been devoted to understanding PDRs.
Comprehensive models of PDRs have been constructed by several groups
(e.g., Black \& Dalgarno 1977, van Dishoeck \& Black 1986, 1988,
Tielens \& Hollenbach 1985a, Sternberg \& Dalgarno 1989,
le Bourlot et al.\ 1993, Kaufman et al. 1999)
by solving the full chemistry and heating-cooling balance in a
self-consistent way.
In PDRs with $n \leq 10^5$ \cmv, the most important heating process for gas
is photoelectric heating---electrons ejected from dust particles
by FUV photons heat the gas through collisions.
For dust, direct absorption of FUV photons
is the primary heating mechanism.
Far-infrared continuum emission is the major cooling process
for dust, and the \fir\ lines of \CII\
$^2$P$_{3/2}\rightarrow ^2$P$_{1/2}$ at 157.7409 \micron\
(hereafter \CII) and
\OI\ $^3$P$_1\rightarrow ^3$P$_2$ at 63.183705 \micron\
(hereafter \OI) are the most important for gas.  Therefore, the \CII\ and
\OI\ lines, along with \hh\ rotational emission, are the most
important tracers of PDRs.

Most previous work has focused on bright PDRs very close to hot OB stars, e.g.,
the Orion bar (Tielens \& Hollenbach 1985b; Jansen et al.\ 1995;
Hogerheijde et al.\ 1995; Tauber et al.\ 1994),
the NGC 2023 PDR (Steiman-Cameron et al. 1997, Draine \& Bertoldi 1996),
and the S140 PDR (Emery et al. 1996; Timmermann et al.\ 1996;
Spaans \& van Dishoeck 1997).
Other recent studies include those by Liseau et al. (1999), and
the field has been reviewed by Draine \& Bertoldi (1999) and by
Hollenbach \& Tielens (1999).
These regions have a \fuv\ intensity of \go\ $> 10^3$
and a density higher than 10$^4$ \cmv, where \go\ is the enhancement
factor relative to the standard interstellar radiation field as given
by Habing (1968).

There has been very little exploration of the physics of PDRs with modest
\fuv\ fields and densities, conditions likely to prevail over most of
the surface of molecular clouds in our Galaxy.
Federman et al. (1995) and van Dishoeck \& Black (1988) have studied PDRs
in diffuse clouds ($A_V \sim 1$ mag) and translucent clouds
( $n< 1000$ \cmv, \go\ $< 17$ and $A_V < 5$ mag).
Regions with high densities and moderate UV fields ( $n< 5\ee4$ \cmv,
$\go\ = 900$) have also been studied in some detail (Jansen et al. 1995),
and Kemper et al. (1999) have used \submm\ and \fir\ observations to
probe a reflection nebula with $n \sim 5000$ \cmv\ and $ \go \sim 200$.
In this paper, we explore the critical intermediate
regime where $n \equiv\ n({\rm H}) + 2 n({\rm H_2}) \sim 500-5000$ \cmv\ 
and $\go \sim 10-60$.
The Infrared Space Observatory (ISO\footnote{ISO is an ESA project
with instruments funded by ESA Member States (especially the PI countries:
France, Germany, the Netherlands and the United Kingdom), with the
participation of ISAS and NASA.}) provided a unique opportunity
to observe low-brightness extended \CII, \OI, and \hh.
We used this capability to study the intermediate regime.

It is also important to understand the role of regions with modest extinction
and column density in star formation.
Regions like those we are studying include most of the mass in the
interstellar medium (Hollenbach \& Tielens 1995),
but their role in star formation is poorly known.
Based on the Jeans criterion,
most molecular clouds in the Galaxy are not sufficiently supported by
thermal energy and therefore should collapse under gravity to
form stars.  Such widespread collapse, however, would lead to a Galactic star
formation rate hundreds of times
higher than what has been observed (Zuckerman \& Palmer 1974, Evans 1991).
The observed low star formation rate seems to indicate that most parts of
most molecular clouds are ``sterile'' (Evans 1999).
Magnetic fields and turbulence are generally considered to play
an important role in supporting molecular clouds and preventing or slowing
collapse.  However, there is no widely accepted theory
on the mechanism of magnetic and turbulent support of molecular clouds.
Recently, Elmegreen (2000) has argued that star formation does in fact happen
within a few cloud crossing times, removing the need for cloud support.
Pringle, Allen, \& Lubow (2001) have pointed out that such a picture strains
methods of cloud formation, and they conclude that visible clouds would have to
form out of ``dark" molecular matter.  These scenarios beg the question:
what prevents star formation in the great majority ($\sim 98$\%) of molecular
gas?

McKee (1989; Bertoldi \& McKee 1996, 1997) proposed a mechanism
of photoionization-regulated star formation to explain the low star formation
rate in the Galaxy by combining two ideas: first, that magnetic fields support
molecular clouds; and second, that PDRs occupy a large fraction of molecular
gas.
The timescale for ambipolar diffusion is proportional to the ionization
fraction ($\xe = n(e)/n$) and the FUV photons
and cosmic rays are the sources of ionization. Most molecular
gas is sterile because it resides in a PDR, where the ionization is high enough
to slow ambipolar diffusion. Only highly shielded regions are likely
to form stars. In addition,
newly formed stars inject energy into the clouds, replenishing
turbulence and slowing star formation in the rest of the cloud.
In this picture, molecular clouds reach dynamic equilibrium
when $\AV \sim 8$ mag.
By focusing on a peripheral region, we test the conditions in
these regions, which should not be forming stars according to the theory.

\begin{figure}[hbt!]
\includegraphics{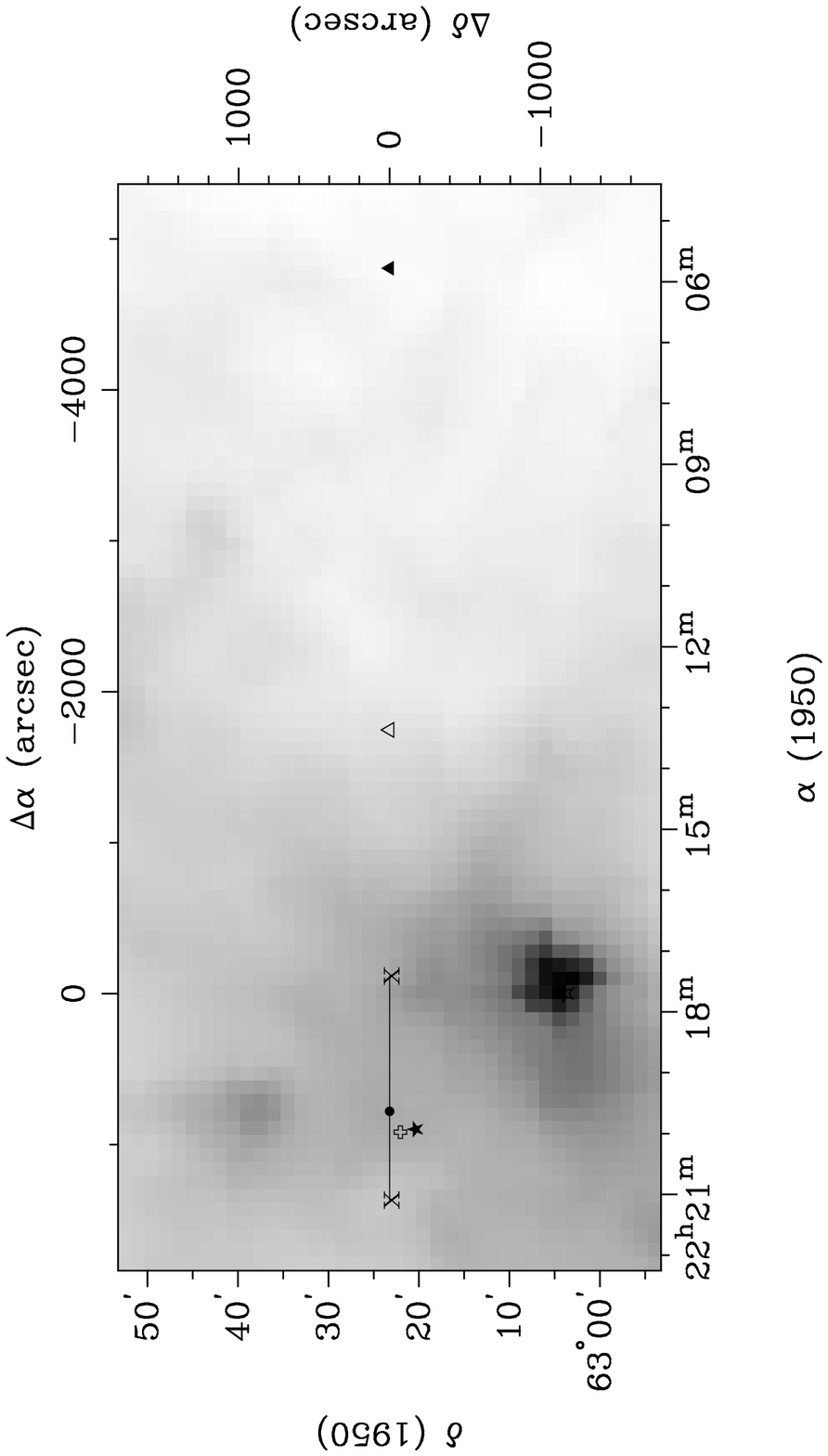}
\vskip 6.1in
\caption{
The S140 cut is marked as the line between two ``X'' symbols
on the IRAS 100 $\micron$  map of the region.  
Darker color indicates stronger 100 $\micron$ emission.
The open triangle is the CSO
Off position, and the solid triangle marks the ISO Off.
The open cross and the solid star are the centers of NH$_3$ (1,1)
cores F and F$'$ identified by Tafalla et al.\ (1993).  The F$'$ core has
the same velocity ($V_{\rm LSR} \sim -7.5$ \kms) as the dense core at IRS1,
while the F core is at a different velocity, $V_{\rm LSR} \sim -10.5$ \kms.
The star marks the position of an IRAS source associated with the F core.
The center of the ionization front is roughly located at 
$22^h$ $17^m$ $30^s$, $63\degree$ $01\arcmin$; 
the exciting star for the ionization front is at 
$22^h$ $16^m$ $50^s$, $62\degree$ $58\arcmin$. 
At the distance of S140, 1000\arcsec\ corresponds to 4.4 pc.
\label{s140-iso}}
\end{figure}

We have chosen a 1-D positional cut---we call it the S140 cut---in the
peripheral region of the molecular cloud L1204 (Figure~\ref{s140-iso})
for our PDR and chemistry study.
S140, at distance of $\sim\! 910$ pc (Crampton \& Fisher 1974),
is an \HII\ region associated with the molecular cloud L1204.
In the rest of the paper, we will use S140 to refer to the whole cloud.
The B0V star HD 211880 illuminates this cloud from the southwest side
to create the visible \HII\ region, an ionization front, and a bright PDR
that has been studied extensively. For a picture that shows the
CO emission in relation to the ionization front and exciting star, see
Fig. 1 of Zhou et al. (1994).
Northeast of the bright PDR lies the
most prominent dense core (hereafter referred to as the S140 dense core)
associated with the cloud.
Numerous studies have been carried out on this cloud, e.g.,
Blair et al. (1978),
Tafalla, Bachiller, \& Martin-P\'{\i}ntado (1993),
Plume, Jaffe, \& Keene (1994), Emery et al. (1996), and Timmermann et al.
(1996).
The FUV radiation at the ionization front provided by HD 211880 is
about 150 times Habing's mean interstellar radiation field
(Keene et al.\ 1985; Habing 1968).
The S140 cut, specified in Table~\ref{isopos},
is far enough ($\sim 20 \arcmin$) 
from HD211880 to have a low FUV radiation field
($\go \sim 15 - 60$, \S \ref{sec4:fuv}, \ref{sec4:kaufman}) and modest
visual extinction ($A_V \sim  15$, \S \ref{sec4:density}).

\section{OBSERVATIONS \& DATA REDUCTION}  \label{sec4:obs}

We have observed the S140 cut with
the Caltech Submillimeter Observatory
(CSO\footnote{CSO is operated by the California Institute of
Technology under funding from the National Science Foundation, contract
96-15025.}) and ISO.
{}From the CSO, we collected data on transitions of \CO\ and other CO isotopes,
the neutral carbon [C I] $^3$P$_2\rightarrow ^3$P$_1$ transition
(hereafter \CI), and the \HCOPTHR\ line (see details in \S~\ref{sec4:cso}).
We use these lines to determine column density of the inner part of
the cloud and to constrain the density.
We used ISO to observe the \CII\ and \OI\ lines along the cut (see details in
\S~\ref{sec4:iso}),
which will help us to determine the \go\ and density
of the cloud, as well as the incident FUV radiation intensity on the clouds.
Finally, we observed rotational transitions of \hh\ with ISO at selected
positions (\S~\ref{sec4:iso}).

Different off-cloud positions have been used
for the ISO and CSO observations, based on the
IRAS 100 \micron\ map and the CO emission (Table~\ref{isopos}).  The ISO
Off position had to be chosen farther away from the S140 cut than the
CSO Off position, because of the greater extent of the \fir\ continuum.
The CSO Off position was used in position-switched observations,
but the ISO Off position was observed in the same way as
the on-cloud positions but with longer integrations, in order to
measure the \CII\ and \OI\ lines from the general background
and establish appropriate baseline intensities.

\subsection{CSO Observations}  \label{sec4:cso}

Various molecular lines and the \CI\ line
were observed using the CSO, with
parameters listed in Table~\ref{cso-obs}.
Most of the data were taken with a re-imaging device
installed on the telescope so that the effective aperture was 
1--2 m, depending on the frequency and the run (Plume \& Jaffe 1995).
The beam size with the re-imager installed, listed
in Table~\ref{cso-obs} for each run, was 2\arcmin--3\arcmin\
and is referred to as the ``big beam''.
Beam sizes of the big beam in Table~\ref{cso-obs}
are accurate to better than 10\%.
Since our object is extended, the exact beam size is
not critical.  The \CI\ line was observed with the full dish
(when the re-imager was unavailable), in the mode of ``On-The-Fly Mapping'';
then the data were smoothed to 120\arcsec\ to match the big beam.
The pointing and the main-beam efficiency were obtained by observing
the available planets and the Moon.  For the full dish observations,
the pointing uncertainty was less than 5$\arcsec$; with the
re-imager, it was no more than 30$\arcsec$.

The CSO facility SIS receivers were used for all observations.  The backend
was a 1024-channel acoustic-optical spectrometer, with a bandwidth of 50 MHz.
The typical single-sideband system temperature was 300 K for the 230 GHz band,
1300 K for the 345 GHz band, and 4000 K for the 490 GHz band.  Observations
were done with position switching between the on- and the off-source
positions.  The data are presented on the T$_{\rm A}^*$ scale.
The rms noise is not uniform for the CSO data,
because of different integration times, weather conditions, and instrument
performance.
At some positions, for some less abundant species,
data were not taken because of extremely weak emission predicted from the
more abundant isotopes and the usual abundance ratios.
Repeated observations showed that uncertainties of the results are
less than 30\%.  

\subsection{ISO Observations}  \label{sec4:iso}

Parameters for the ISO observations are in Table 
\ref{iso-lines}.  Information
about ISO can be found in Kessler et al.\ (1996).  We used
the Long-Wavelength Spectrometer (LWS; Clegg et al.\ 1996;
Swinyard et al.\ 1996) to observe the \CII\ and \OI\ lines in
the LWS02 mode at 16 positions along the S140 cut.
For each line at every
position, multiple scans were obtained; each scan had 19 samples
covering a width of 5 spectral resolution elements.
Each sample had an integration time of 0.4 sec.
Particle hits caused a number of bad scans, resulting in a significant
reduction of usable data.  Three selected positions from the S140 cut
were observed in the \CII\ and \OI\ lines with longer integrations.
The ISO Off position of S140 was also observed in the \CII\ and the \OI\ 
lines.
The data presented here have been processed through the Off-Line Processor
Version 8.7 (OLP 8.7),
and the line fluxes were extracted
by using the ISO Spectral Analysis Package Version 1.6a (ISAP 1.6a),
calculated under the assumption that the
sources were point sources. Processing consisted of removing
bad scans and obvious glitches and, in a few cases, DC offsetting
the continuum to be closer to that of the other scans. The points
were then averaged together using $2.5 \sigma$ clipping and fitted
with a line and baseline fit. The quoted errors are those given
by the fit. Because the standard calibration assumes a point source
and, in the region we mapped, the line emission is more extended than the 
beam, we multiplied these fits by 0.84 for the 63 um \OI\ line and 0.59
for the 157 um \CII\ line, using Table 4.10 of Gry et al. (2001).
We then converted to surface brightness, using
beam solid angles listed in Table 4.14 of Gry et al. (2001).
The relevant detectors are SW3 (\OI) and LW4 (\CII), resulting in beam
solid angles of 1.30\ee{-7} for \OI\ and 6.65\ee{-8} for \CII.
The \OI\ solid angle is smaller than the nominal \iso\ beamsize 
of 1\farcm6 but considerably larger than the diffraction limit.
The \CII\ beamsize is slightly larger than the diffraction limit 
at 158 $\mu$m.
The point source fluxes were multiplied by the following factors
to get surface brightness in \icgs:
\OI\  6.46\ee{13} ; \CII\  8.87\ee{13}.
In addition to our own cut through the cloud, we have reprocessed the LWS
data taken toward the bright PDR near the far-infrared peak in S140 itself,
originally taken for the program of Emery et al. (1996, TDT 09101820 in the
ISO Data Archive) and for
another program (TDT's 82301120, 122, and 123) using the same corrections
as for the data from our own program.

Because the LWS has ten detectors,
we also obtained undersampled \fir\ continuum scans covering from about
43 \micron\  to 175 \micron\ from the detectors that were not on the
lines.  However, because of dark current uncertainty
for the bands shorter than 100 \micron\ and the fringing problem
for the bands longer than 100 \micron\ (Swinyard et al.\ 1996), we could
not make accurate measurements of the \fir\ continuum from the ISO scans.
Therefore in this work, we will use the IRAS measurements to constrain
the \fir\ emission.

The low-lying  pure-rotational H$_{2}$  $J=2\to0$ S(0) line  at 28.218 $\mu$m
and the $J=3\to1$ S(1) line at 17.035 $\mu$m were observed with the ISO-SWS
in the AOT02  mode (de Graauw et al.\ 1996). Typical integration times were
$\sim$100~s per line, in which the 12 detectors were  scanned once  over the
28.05--28.40 and  16.96--17.11 $\mu$m ranges  around the lines. The  $J=5\to
3$ S(3)  9.66 $\mu$m and $J=7\to 5$ S(5)  6.91 $\mu$m lines were measured  in
parallel with the S(0) and  S(1) lines, respectively.  
The spectral resolving power for point sources  is 2000 (150 km
s$^{-1}$) at
28 $\mu$m, 2400 (125 km s$^{-1}$) at 17 $\mu$m, 2280 (130 km s$^{-1}$) at 9.7
$\mu$m  and  1550 (195  km  s$^{-1}$)  at  6.9 $\mu$m.   The  SWS apertures
are $20''\times  27''$ at S(0), $14'' \times  27''$ at S(1), and $14''\times
20''$ at the S(3) and S(5) lines. Since the SWS beam is considerably smaller
than the LWS beam, three positions separated by 30$''$ in RA were observed for
each of the three LWS positions and averaged to produce an intensity
for comparison to the LWS data.

The H$_2$ lines from translucent clouds are predicted to be weak and close to
the sensitivity limit of the SWS instrument. At this level, noise induced by
charged-particle impacts  on the detectors plays a large role.  By good
fortune, these particle hits were exceptionally low during the orbit in which
the H$_2$ data for S140 were taken so that the quality of the data ($\sim$0.2
Jy rms at 28 $\mu$m) is comparable to that commonly obtained in much longer
integration times. The H$_2$ S(0) and S(1) lines were detected with the 
standard pipeline
reduction, but special  software designed to  handle weak  signals in
combination  with  the  standard  Interactive  Analysis Package was used for
improvements in the signal-to-noise ratio. This technique also yields a 
possible detection of the S(3) line at one position. 
The details  and justification  of
the  methods used  in the software are described  by  Valentijn \&  Thi
(2000) and Thi et al.\ (2001). Because the mid-infrared continuum emission is
weak, $<$ 3 Jy, the data do not suffer from fringing effects caused by an
inadequate responsivity function correction.  However, the uncertainties
are dominated by the removal of the fluctuating dark current (see Leech et
al. 2001), leading us
to adopt a calibration uncertainty of $\sim 30$\%, which is propagated 
in the analysis.

\section{OBSERVATIONAL RESULTS} \label{sec4:res}

\begin{figure}[hbt!]
\includegraphics{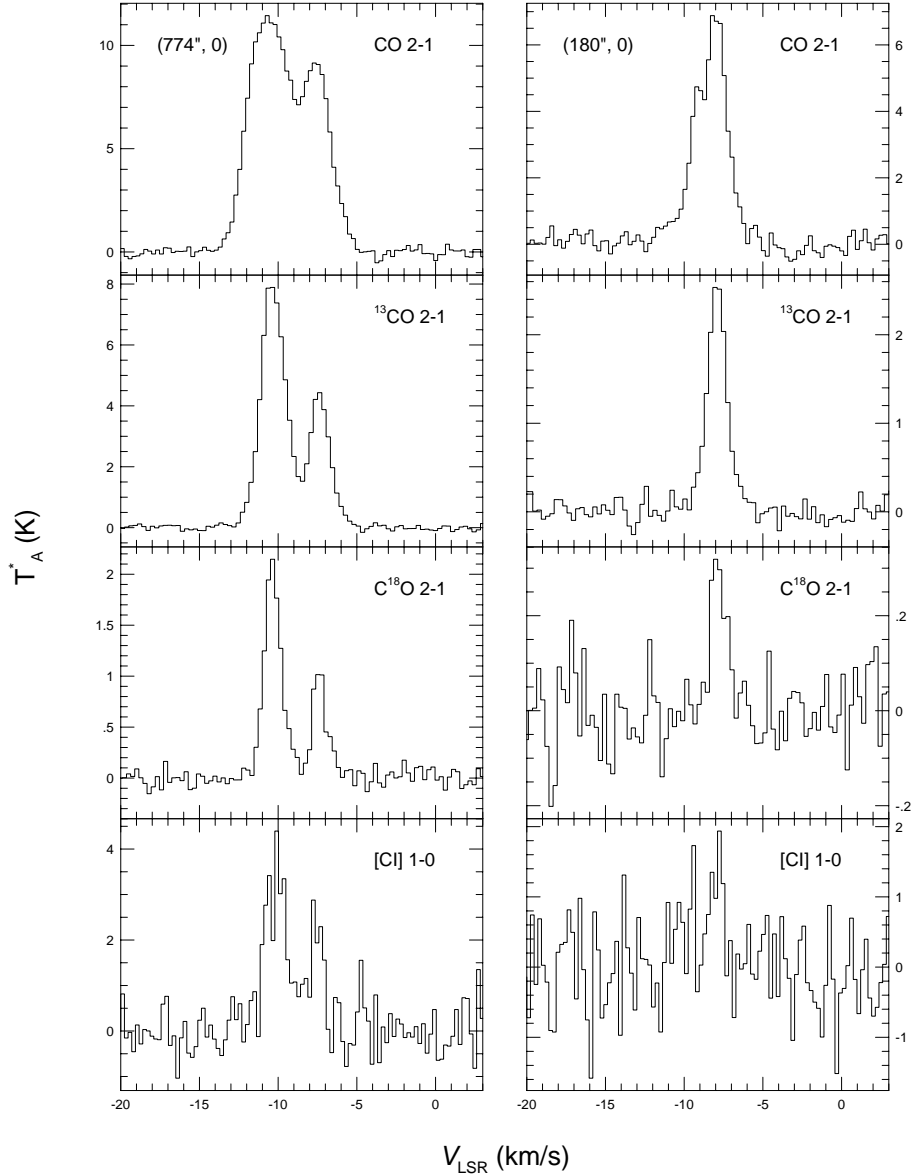}
\vskip 6.60in
\caption{
Spectra of $J\!=2\!\rightarrow\!1$ lines of \CO, \CCO, \COO, and
\CI\ 492 GHz line at two selected positions.
At the position (180$\arcsec$, 0), the line velocity
($\sim -7.5$ km/s) is the same as that at the dense core at IRS1;
thus the lines here are from the cloud extension of the dense core.
At the position (774$\arcsec$, 0) two velocity components show up, one being
the same as that of the dense core, another at about $-10.2$ km/s.
The second component is from a distinct clump, as evidenced
by the study of Tafalla et al.\ 1993
         \label{csolines-sp}}

\end{figure}

\begin{figure}[hbt!]
\includegraphics{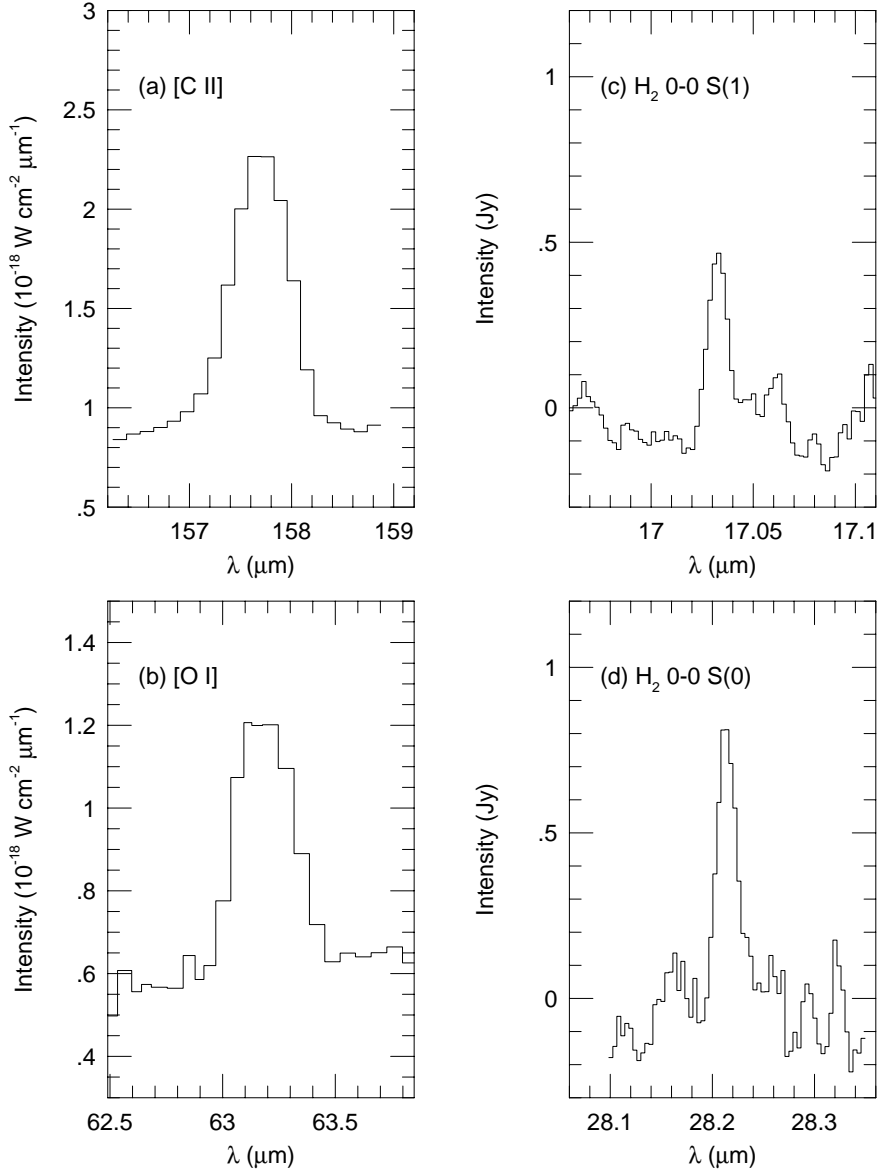}
\vskip 6.8in
\caption[/Figures/isolines.ps]
{ISO spectra averaged along the S140 cut.
None of the lines are spectrally resolved.
The \CII\ and \OI\ spectra (panels a and b) are the averages of the 16 
positions
along the S140 cut, as listed in Table 6.
The two H$_2$ spectra (panels c and d) are averaged over the 3 LWS
positions observed with SWS (Table 7).
\label{isolines-sp}}

\end{figure}

Observational results from the CSO and ISO are presented in
Tables~\ref{s140cso-res}--\ref{s140sws-res}.
Spectra of \COTWO, \CCOTWO, \COOTWO, and the \CI\ line
at two selected positions are shown in Figure~\ref{csolines-sp}.
Figure~\ref{isolines-sp} shows the spectra of the \CII\ and the \OI\ lines,
averaged over the whole S140 cut, along with the \hh\ S(0) and S(1) lines
averaged over the three LWS positions where SWS observations were made.

Two velocity components of $V_{\rm LSR} \sim -7.5$ \kms\ and
$V_{\rm LSR} \sim -10.5$ \kms\ are clearly seen in the CSO data
along the S140 cut.  In
Table~\ref{s140cso-res}, results for each component are given when applicable.
The line widths (FWHM) of the \COOTWO\ transition of both
components are around 1.3 \kms, implying a Doppler parameter, $b = 0.78$,
where $b$ is the $1/e$ half-width of the line.
The first component is very close to the velocity observed at the S140
dense core.  Therefore the molecular gas
with this velocity is likely to be the cloud extension of the S140 dense core.
The second component, strongly peaking around (774\arcsec, 0)
(Table~\ref{s140cso-res}) for the \CI\ line and the lines of \CO\ and
other CO isotopes, is the same as that of core F
identified in CS\,$J\!=\!1\!\rightarrow\!0$ and NH$_3$ (1,1) and (2,2)
by Tafalla et al.\ (1993).
This dense core, apparently active in star formation with an associated
IRAS source and outflow, is centered at (954\arcsec, $-60$\arcsec)
(relative to the reference point of this work in Table~\ref{isopos})
and extends about $2\arcmin \times 2\arcmin$ in NH$_3$ (1,1).
We distinguish the emission in this component from the cloud
by referring to it as core F; the first component is referred to as
the cloud.  We define the mean emission in core F as the average of
the 5 positions with offsets in $\alpha$ between 576\arcsec\ and 972\arcsec,
using only the $-10.5$ \kms\ component. These positions
include all with obvious effects from core F.
The mean emission from the cloud is the mean of emission from all
positions in the $-7.5$ \kms\ component added to the mean emission from
the $-10.5$ \kms\ component outside that range of $\alpha$.
We add the two components for comparison to the ISO
observations, which are not spectrally resolved.

Table~\ref{s140-hcop} presents results of the \HCOPTHR\ line
at selected positions.
At ($774\arcsec$, 0), we detected a 0.2 K line of \HCOPTHR\ with the
re-imager.  Following this detection, the
full dish was used to map an area of $300\arcsec \times 300\arcsec$
centered at ($774\arcsec$, 0).
The small-beam map revealed that the emission is clumpy.  The strongest
clump is close to the peak of NH$_3$ (1,1) detected by Tafalla et al.\
(1993) (see \ Figure~\ref{s140-iso}).  In Table~\ref{s140-hcop}
we list the small-beam results at the peak position.
Big-beam observations of the \HCOPTHR\ line
at other positions did not result in any detection up to the
standard deviations given in Table~\ref{s140-hcop}.

Measurements of the \CII\ and \OI\ lines are given in
Table~\ref{s140lws-res}, as is the ratio, \OI/\CII, and
the IRAS continuum intensities at 60 \micron\ and 100 \micron.
Line measurements from
the combined data set of all 16 positions along the S140 cut (Avg(16-pos)),
at the S140 PDR peak position, and at the ISO off position are also given
in the table. The 1$\sigma$ uncertainties come from the
Gaussian fits performed by ISAP. These uncertainties are appropriate
in evaluating the reliability of spatial variations in line strengths
and line ratios.
In comparisons of the absolute values of these quantities to theoretical
models,
the overall calibration uncertainty  provides a more useful
assessment of the significance of the results.
The overall uncertainty results from a combination of model uncertainties for
the primary calibrator (Uranus), variations in the detector dark current, and
uncertainties in the value of the system solid angle at each wavelength
(Swinyard et al. 1998). Considering the various correction factors and their
uncertainties, we adopt a value of 30\% for this overall uncertainty.

\begin{figure}
\includegraphics{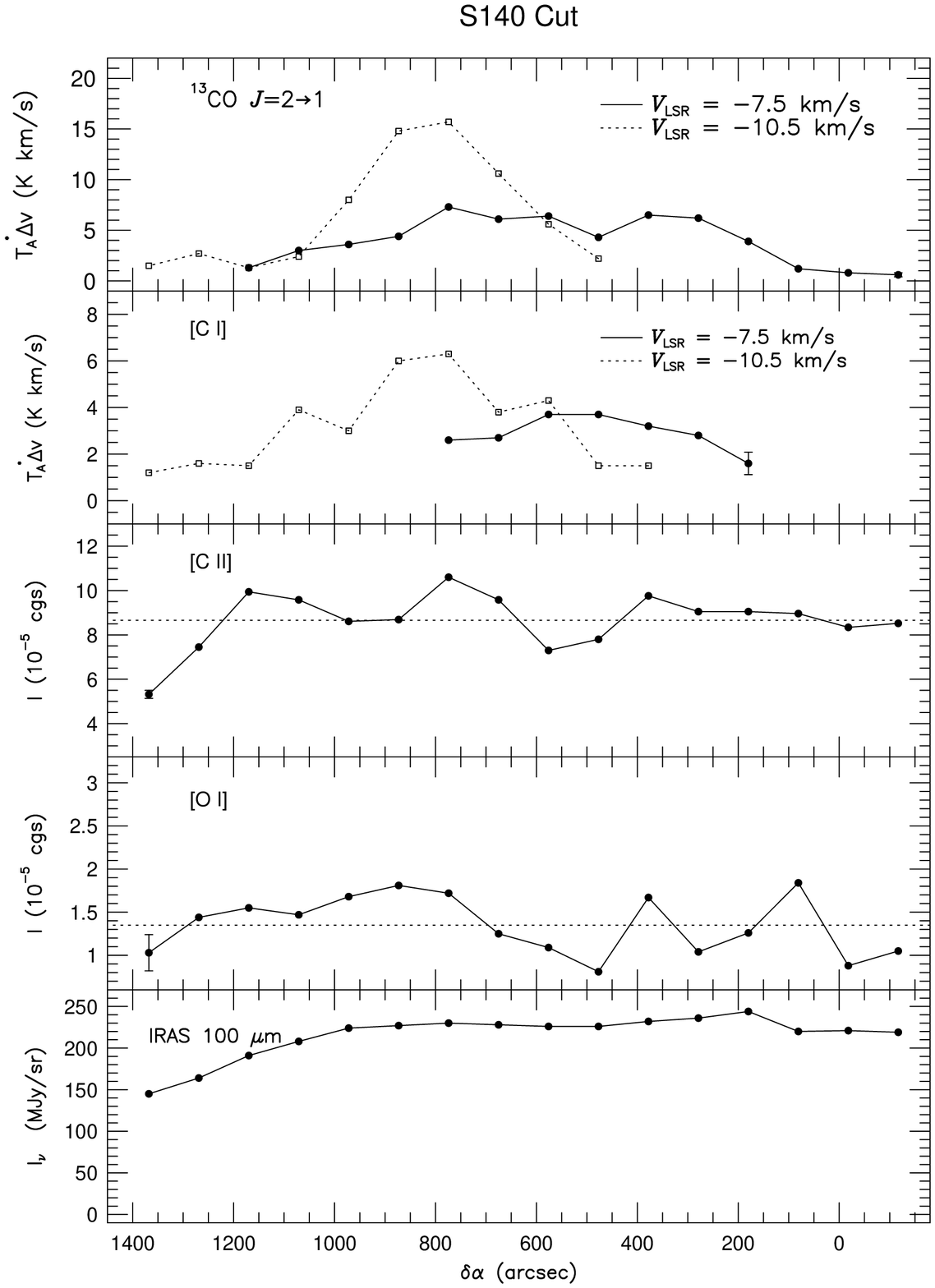}
\vskip 7.1in
\caption{
Spatial variation along the cut of the \CCOTWO, \CI, \CII, \OI lines
         and 100 \micron\ continuum. For the \CCOTWO\ and \CI\ lines,
both velocity components are shown with different line types.
The dotted lines for the panels of the \CII\ and \OI\ lines are the
averaged values of the 16 positions (Avg(16-pos). Representative
error bars are shown on one or the other of the end points; these
error bars do not include calibration uncertainties as they do not
affect the {\it shape} of the distribution.
                  \label{s140cut-lines}}

\end{figure}

Figure~\ref{s140cut-lines} shows the positional variations along the S140
cut of the \CCOTWO\ and \CI\ lines (of two velocity components separately), 
the \CII\ and \OI\ lines, and the intensity of the IRAS 100 \micron\
continuum.  
The dotted lines for the panels of the \CII\ and \OI\ lines mark
the line measurements of all the data combined of the 16 positions
(Avg(16-pos) in Table~\ref{s140lws-res}).
Clearly, the \CII\ and \OI\ lines and the 100 \micron\ continuum are quite
flat over most of the S140 cut, while the \CCOTWO\ line displays
more than a factor of 10 variation in both velocity components, 
and the \CI\ varies by a factor of 4.
The \COTWO, \CCOTHR, and \COOTWO\
show strong spatial variation similar to the \CCOTWO\ line.
In contrast, the average \CII\ and \OI\ fluxes of the 16 positions
(Avg(16-pos)) agree with the measurements at each individual position
within 40\%.
Thus, core F is simply not apparent at all in the \CII, \OI, or \fir\
emission.  Therefore, we will use the \CII\ and  \OI\ 
data averaged over all 16 positions as representative of the cloud.
Core F is also not noticeable in the more limited \hh\ data (Table 
\ref{s140sws-res}).

The average \CII\ and \OI\ intensities are
$8.66\times 10^{-5}$ and $1.35\times 10^{-5}$
erg~s$^{-1}$~cm$^{-2}$~sr$^{-1}$ respectively, and the average line ratio
\OI/\CII\ is $0.16\pm0.04$.
If we subtract the emission at the ISO-OFF position from the mean
values, the intensities for \CII\ and \OI\ are 6.97\ee{-5} and
0.89\ee{-5} \icgs, with a ratio of \OI/\CII\ of 0.13.
These numbers are very different from the S140 bright PDR,
where the ratio of \OI/\CII\ is $1.55 \pm 0.10$, based on our
reduction of the archival data, using an analysis like that used
for our own data.
For the peripheral regions, \CII, rather than \OI, is the primary
coolant.

The intensities of the \hh\ lines are roughly constant at
$\sim \eten{-5}$ erg~s$^{-1}$~cm$^{-2}$~sr$^{-1}$ at the three positions
(Table \ref{s140sws-res}). In addition, the excitation temperature between
$J = 3$ and $J = 2$ is also constant within uncertainties at $\sim 100$ K.
The constancy of intensity and temperature is rather remarkable because
one position
is toward core F and the other two are positions near the ends of the
cut, without emission from lines tracing molecular gas of moderate density
(cf. Table \ref{s140cso-res}). Clearly, the \hh\ emission has the
same pattern of constancy as the \CII\ and \OI\ lines and traces the
gas in the PDR.

The \hh\ S(3) line is only tentatively detected at one position, with
an excitation temperature between $J=5$ and $J=3$ of $\sim 350$ K.
The non-detections at the other positions give a comparable upper limit
on the temperature.

\section{MODELING AND DISCUSSION} \label{sec4:model}

In this section, we will attempt to reproduce the observations
by using both published models and our own PDR models.  
The most important free
parameters in the modeling are the FUV radiation intensity, the total
extinction, and the volume density.
Therefore, we will first constrain the FUV intensity (\S~\ref{sec4:fuv})
and the extinction and density (\S~\ref{sec4:density}).
Then we will compare the predicted \CII\ and \OI\ line intensities
with the ISO observations in \S~\ref{sec4:kaufman}, \ref{hst}.

\subsection{The FUV Radiation Field Estimated from the Far-infrared Continuum}
   \label{sec4:fuv}

Because dust grains absorb most of the FUV photons incident on the cloud
and most of that energy is re-emitted in the form of \fir\ continuum,
the intensity of the FUV radiation can be inferred from the \fir\ continuum.
Hollenbach et al.\ (1991) presented model predictions of the \fir\ continuum
intensities at 100 \micron\ and 60 \micron\ for a range of incident
FUV intensity.

{}From Table~\ref{s140lws-res} and Figure~\ref{s140cut-lines},
the IRAS 100 \micron\ and 60 \micron\
intensities are quite uniform along most of the S140 cut
except at Positions 15--16 for 60 \micron\ and the Positions 13--16
for 100 \micron, where the intensities are lower.
Therefore we use the intensities averaged over positions 1-13 and the
intensity at position 16 to estimate a range of values for the
ultraviolet intensity.
The continuum intensities at the ISO Off position have been subtracted
to yield $I(60) = 30 - 50$ MJy sr$^{-1}$ and $I(100) = 100 - 180$ MJy
sr$^{-1}$.
Assuming a $\lambda^{-1}$ dust emissivity, the total \fir\ intensity is about
$6\times 10^{-3}$ ergs s$^{-1}$ cm$^{-2}$ sr$^{-1}$.

These intensities and a $\lambda^{-1}$ dust emissivity yield a dust temperature
of $(28\pm 2)$ K, near the constant value of 27 K reached for $\go < 160$,
when effects of transient heating of small grains/PAHs by single photons
are included (Hollenbach \& Tielens \ 1995).
Therefore, the 60 \micron\ continuum intensity in this region
is probably strongly affected by small grains.
Based on Fig. 18 of Hollenbach et al. (1991),
we used the 100 \micron\ intensity to estimate the strength of the
ultraviolet field as $\go = 40 - 60$.
As noted by Kaufman et al. (1999), these values assume that the FUV
impinges only on one surface of the cloud. If the cloud is heated from
both sides, then the values of \go\ appropriate to one surface would
be 20 -- 30. The assumption here is that the cloud is optically thin to the
\fir\ continuum.
Given the uncertainties in the fraction of the grain heating caused by
FUV photons, we estimate \go\ in the range, 15--60.
The enhancement factor of the radiation field $I_{UV}$ with respect to the 
standard field given by Draine (1978) is 1.71 times smaller 
than \go, $I_{UV} = \go/1.71$ (Draine \& Bertoldi 1996).

\subsection{Column Density and Density Regime}  \label{sec4:density}

The mean integrated intensity of \COOTWO\
for the cloud component is 0.54 K \kms.  From Table~\ref{s140cso-res},
the typical excitation temperature of \CO\ is about 10 K.
Assuming that the \COOTWO\ line is optically thin and in LTE,
we calculate analytically
the total column density of \COO\ to be $\leq 1.3\times 10^{14}$ cm$^{-2}$.
The empirical relations of Frerking, Langer, \& Wilson (1982) on $\rho$ Oph
yield a visual extinction, $A_V = 3.4$ mag, and
a total column density of H nuclei of $5\times 10^{21}$ cm$^{-2}$.
Assuming the cloud is spherical and the angular size of the cloud
along the S140 cut is roughly 15\arcmin, the linear size of the depth
of the cloud would be $1.2\times 10^{19}$ cm (4 pc) at a distance of 910 pc.
Therefore, the average density along the S140 cut would be 800 \cmv.
The assumption of LTE is not valid at such a low density.
Models of excitation, including trapping with
LVG codes, indicate that the \COO\ populations at 
such low densities are far from LTE.
A more self-consistent solution is a density around 2000 \cmv\ and a column
density of \COO\ of 2.0\ee{15} \cma, yielding $A_V = 16$.
The ratio of $J=2-1$ to $J=3-2$ lines of \CCO\ is roughly consistent
with a density of 2000 \cmv, and the non-detection of \HCOP\ $J = 3-2$
outside core F limits the density to $n < 10^4$ \cmv\ for typical
abundances in translucent clouds and $\tk = 10-20$ K.
Similar considerations applied to the position of core F (774,0)
lead to estimates of $A_V = 25 $ mag and, using a size of 3\arcmin,
a mean density of \eten4 \cmv.

Considering the uncertainties in these estimates, the likely 
uncertainties in column density and density are factors of 2 and 3,
respectively. A direct comparison between PDR models and our data
will be given in \S \ref{hst}, which supports these simple estimates.

\subsection{Comparison to Published PDR Models}   \label{sec4:kaufman}

We first compare our results to the published grid of PDR models by 
Kaufman et al. (1999). These calculations incorporated new collision rates for
fine-structure lines and \hh, new PAH heating and chemistry, and lower
gas-phase abundances for oxygen and carbon (Savage \& Sembach
1996). They include regions of low FUV and low density, which seem to
be appropriate for our situation. The calculations employed a turbulent
broadening of 1.5 \kms, about twice what we infer, but Kaufman
et al. argue that the results are not too sensitive to this parameter.

We first consider the constraints from the two atomic species, \CII\
and \OI. From Fig. 4 of Kaufman et al., the mean ratio of \OI/\CII\ 
of $0.16\pm 0.4$ constrains $\go < 20$.
This is at the low end of the range
inferred in \S \ref{sec4:fuv}, even if we assume the cloud is heated from both 
sides.
At this \go, the only solution for density is $n \sim 300$, 
lower than the values inferred in \S \ref{sec4:density}.
For lower \go, the solutions for $n$ bifurcate into
lower and higher values. For example, at $\go = 10$, there are solutions
for log $n = 2.0$ or 3.3.
The latter are more consistent with the LVG modeling of \COOTWO.
The \CII\ emission by itself (Fig. 3 of Kaufman et al.),
is less diagnostic, but roughly consistent with the ranges implied by
the ratio.

Are these low values of \go\ consistent with the \fir\ emission?
They are substantially below the estimates from \S \ref{sec4:fuv}.
Our ratio of intensities of the sum of the \OI\
and \CII\ lines to the total \fir\ intensity, 1.7\ee{-2}, is close to the
highest values found by Kaufman et al. (1999).
In fact, the ratio of line to continuum
of the cloud in S140 may exceed the maximum in their models, if indeed
the cloud is heated from both sides.

What about the \CI\ data? The mean for the cloud, summing the two components
as for the \COO\ data, and correcting for efficiency, is
1.3\ee{-6} \icgs. Comparing to Fig. 7 of Kaufman et al., this value
favors very low \go. To get \go\ up to even 15 requires $n \sim 100$.
For the conditions that match the other lines, the \CI\ prediction of Kaufman
et al. is too high by a factor of 3--4.

Given reasonable uncertainties, the mean cloud data require
quite low values of \go\ and $n$, with most likely values of
$\go \sim 15$ and  $n \sim 1000$. These are both lower than
our initial guesses, even if the cloud is heated from both sides. For this
range of conditions, the ``surface temperature" of the PDR, plotted in
Fig. 1 of Kaufman et al., will lie between 100 and 200 K. This is the
maximum temperature that should apply to the region of \hh\ emission,
if collisions dominate the \hh\ excitation.
The excitation temperatures of \hh\ (Table \ref{s140sws-res}) are
consistent with this range of temperatures, but the absolute
intensities are difficult to reproduce (\S \ref{sec4:h2interp}).

Comparison of the Kaufman et al. results with those of other models
in the literature (e.g., Le Bourlot et al.\ 1993, Roueff, private
communication) and our own models (e.g., Jansen et al.\ 1995, Spaans \& van
Dishoeck 1997, Spaans, private communication), shows good agreement for the 
\CII\ intensities but variations of factors of two in the \OI\ intensities
depending on the temperature structure, radiative transfer treatment and
geometry of the source. Thus, the range in inferred \go\ and $n$ could be
somewhat larger, but the conclusion that both are low holds firm.

The region of the core F emission is clearly denser than that of the
cloud, yet core F is not apparent in either \fir\ lines or continuum.
The emission from the \CII\ line may be quite independent of density up
to about $n \sim \eten4$ \cmv, in the allowed range of \go; however, the \OI\
line should increase with density. From Fig. 4 of Kaufman et al. (1999),
the ratio of \OI/\CII\ should increase to about 1 for $n = \eten4$ and
\go\ in the range that fits our data. To keep the ratio below 0.3
would require  $\go \sim 1$. That is, core F would have to be
shielded from the ambient FUV field. Interestingly, the \CI\ emission
does see core F in the $-10.5$ \kms\ line. Either that component is not
completely shielded or the enhanced \CI\ is caused by some internal source
that does not affect the \fir\ line and continuum emission.

\subsection{Comparsion to PDR-Monte Carlo Models} \label{hst}

To check the effects of radiative transport, different Doppler broadening,
and metal abundance at these low values of \go\ and $n$, we ran our
own models for the part of parameter space indicated in the last section.
We used our own PDR code, an updated version of the code described by
Jansen et al. (1995) to calculate abundances of relevant species 
as a function of depth into the cloud. 
We then fed the results into a Monte Carlo code (Hogerheijde
\& van der Tak 2000) to calculate excitation and radiative transport.
Einstein A values of \CII\ and \OI\ were taken from Tielens \&
Hollenbach (1985a) and Galavis et al. (1997), respectively.
In going from the PDR code to a spherical Monte Carlo code, one doubles
the column density, so the PDR models were run for $\AV = 9$, yielding
a total extinction through the cloud of 18 mag. We ran models for Doppler
parameters from 0.5 to 2.0. The integrated intensities were generally
insensitive to this parameter. Two sets of values for metal abundances
were used: normal abundances are those listed by Jansen et al. (1995) in
their Table 2 for the $\zeta$ Oph diffuse cloud;
``low metals" corresponds to decreasing all
abundances other than C, N, and O by a factor of 10. The most important
species is S, which is thought to be quite depleted in molecular clouds
compared to translucent clouds. 

The results showed that the radiative transport of \OI\ is indeed quite
sensitive to the input parameters because it is very subthermally excited
at these low densities, but quite opaque ($\tau \sim 10$ for models that fit 
the data well). At very low values of \go\ and $n$, the \OI\ line 
predicted by the Monte Carlo cloud can be considerably stronger than 
that predicted by the PDR code alone, after accounting for the
doubling of the column density. The differences seem to be partially
in the handling of the radiative transport and partly in the geometry.
Calculations that calculate the temperature self consistently 
in spherical geometry produce even larger effects (M. Spaans, personal 
communication).

Among a grid of models, the best fit was obtained for a model with
$\go = 17$ and $n = 1000$ and low metals. 
This model reproduced the \COO\ line strength, indicating that the
extinction estimate from \S \ref{sec4:density} was about right.
The ratio of \CCOTHR/\CCOTWO\ was about 50\% lower than the observations
in this model, while models with $n = 3000$ gave ratios higher than observed
and \CCOTWO\ lines about a factor of 2 stronger than observed.
The \HCOPTHR\ line produced by
this model is about 1.5 times our RMS noise, still consistent with the 
observations, while the model with $n = 3000$ would produce a 10 $\sigma$
detection.
The \HCOP\ abundance and thus the \HCOPTHR\ line are considerably
enhanced in the case of low metals. Taken together, these results
suggest that $n = 1000$ \cmv\ is about right, but perhaps a bit low. 

The best model gave
a \CII\ line in good agreement with observations and an \OI\ line
about 40\% too strong. Higher values of \go\ or $n$ greatly overproduced
the \OI\ line. These are in rough agreement with Kaufman et al., but the 
\CI\ line predicted by our models is about a factor of 4 weaker than the
line predicted by Kaufman et al., and our prediction agrees with the 
observations.
Kaufman et al. note that their \CI\ lines are stronger than in previous 
calculations
because of inclusion of PAH chemistry in their models. While charge transfer 
from 
PAHs to \CP\ to create \CZ\ is included in our models, the details may be 
different.

We use the results of the model with $\go = 17$ and
$n = 1000$ in the next section.

\section{TESTING PHOTOIONIZATION-REGULATED STAR FORMATION} \label{sec4:mckee}

Figure 5 shows the distribution of electron fraction ($x(e) \equiv n(e)/n$,
with $n = 2n(\hh) + n(H)$) with extinction
from the best fitting models from \S \ref{hst}.  The model with
low metals fits the \COO\ data better and has slightly lower \xe.

The timescale for ambipolar diffusion is:
\[
t_{\rm AD} \simeq 1.6\times 10^{14}  x(e)~~~~{\rm yrs},
\]
where \xe\ is the electron abundance (McKee et al. 1993).
Since the lifetime of molecular clouds is roughly 30 Myrs
(Bash, Green, \& Peters 1977), if $t_{\rm AD}$ exceeds this time,
the region is effectively sterile as long as it remains magnetically
sub-critical: it will be dissipated before it
can form stars. By setting \tad\ to 30 Myr, we find a threshold
ionization fraction for star formation of
$\xe \simeq 2\times 10^{-7}$.
In Figure 5, the threshold $\xe \simeq 2\times 10^{-7}$ is reached
at $A_V = 6$ mag for the low metal model and is not reached
by the end of the high metal model at $\AV = 9$ mag.
A cloud with a column density less than 6--10 mag will have little
well-shielded ($\xe < 2\times 10^{-7}$) gas to form stars.
Thus the ionization fraction in the region we are studying suggests
that little star formation would be expected in the extended cloud,
if photoionization-regulated star formation applies (McKee 1989).

Based on the above analysis, the localized core F along the S140 cut,
with $A_V \sim 25$ mag, has substantial gas that is sufficiently 
well shielded ($\xe < 2\times 10^{-7}$) to allow
star formation, which has been confirmed by the outflow around this
core (Tafalla et al.\ 1993).
This is consistent with McKee's theory of photoionization-regulated
star formation.
In comparison, most of the molecular gas in the extended cloud,
with $A_V \sim 16$ mag,
has a $\xe > 2\times 10^{-7}$ and is unlikely to form stars, if
subjected to FUV radiation from both sides of the cloud.
Infrared studies to look for evidence of star formation in
the peripheral region of S140 (away from core F)
would be able to test this prediction and McKee's theory.

Since the average extinction of molecular clouds in the Galaxy is
about 7.5 mag (Larson 1981),
according to the photoionization-regulated star formation model,
most of the molecular gas will not form stars,
leading to the low observed Galactic star formation rate.
The sterility of much of the molecular gas is supported by
the apparent absence of distributed populations of embedded stars
in NGC 2023 (Li, Evans, \& Lada 1997).  
The extinction in the peripheral region of that cloud,
estimated from \CCOONE\ emission is less than 10 mag, while that
in the core is about 14 mag, just enough for shielding to decrease the
ambipolar diffusion time. More recently, Carpenter (2000) has
used 2MASS data to set upper limits on the distributed population
in several clouds, again consistent with the result that star formation
is largely confined to clusters forming in gas of higher than average 
extinction and density.

Of course, a long $t_{\rm AD}$ is only a necessary condition for the
picture of photoionization-regulated star formation, not a sufficient one.
The clouds must be magnetically subcritical as well. Crutcher (1999)
has summarized the data on this topic, concluding that clouds were
slightly supercritical, but he noted that the uncertainties allowed
the clouds to be subcritical.  Shu et al. (1999) have noted
some corrections that make the clouds even closer to the critical boundary,
and they suggested that observed clouds may be self-selected to be
close to marginally critical. 

\section{THE \hh\ EMISSION} \label{sec4:h2interp}

The H$_2$ pure-rotational lines observed with the ISO-SWS provide a new test
of the PDR models. Early ISO-SWS results have shown that H$_2$ lines up to
$J=7\to 5$ S(5) are readily detected in traditional high density ($> 10^4$
cm$^{-3}$), high \go\ ($>10^3$) PDRs such as S140 at the position of the
interface with the H~II region (Timmermann et al.\ 1996, see review by Wright
2000). Those data give excitation temperatures for $J=3-7$ of 400--700~K,
much higher than expected from pre-ISO PDR models (Bertoldi 1997, Draine \&
Bertoldi 1999). Because of the high densities, the H$_2$ level populations are
thermalized so that the H$_2$ intensities directly trace the temperature
structure in the warm PDR layers. To remedy these discrepancies, Bertoldi
(1997) suggested that either the photoelectric heating efficiency needs to be
increased or that the H$_2$ formation rate on grains needs to be larger at
high temperatures, shifting the H$\to$H$_2$ transition zone closer to the warm
edge. Weingartner \& Draine (1999) have calculated increased photoelectric
heating efficiencies based on an enhanced dust-to-gas ratio in the PDR due to
gas-grain drift, and Draine \& Bertoldi (2000) have shown that such models can
reproduce the H$_2$ observations of NGC 2023. Note that these comparisons
include a factor of 5--10 enhancement of the models due to
limb-brightening/geometry.

In the low-density, low \go\ regions of S140 studied here, the H$_2$
populations should be controlled by ultraviolet pumping rather than
collisions, at least in the outer layers of the cloud. Moreover, enhanced
efficiencies due to gas-grain drift should not play a role, nor should there be
any geometrical effects in this face-on cloud. Thus, the ISO-SWS data should
provide a test of the H$_2$ excitation in a different PDR regime (Thi et al.\
1999). For the PDR parameters derived in the previous sections, $n\approx
1000$ cm$^{-3}$ and $\go \approx 15$, standard low density PDR models such
as those by Hollenbach et al.\ (1991) or le Bourlot et al.\ (1993) and our own
models give H$_2$ excitation temperatures derived from the S(0) and S(1) lines
of $\sim$100--200~K, consistent with our ISO data. However, these models
underproduce the absolute H$_2$ intensities by at least an order of magnitude.
The same conclusion was reached by Habart et al.\ (2000) and Kemper et al.\
(1999) for somewhat denser PDRs exposed to low \go. Habart et al.\ proposed
to increase the H$_2$ formation rate on grains by more than a factor of 5
compared with the standard H$_2$ formation rate of $\sim 3\times 10^{-17}$
cm$^3$ s$^{-1}$ at 100 K (Jura 1975) to reproduce their observations.

If the H$_2$ formation rate in our S140 models is increased by a similar
factor of 5, the comparison with observations is improved considerably,
although the models still fall short by a factor of a few. Contributions from
the far-side of the cloud could help to further reduce the discrepancies. Such
an arbitrary, large increase of the H$_2$ formation rate is inconsistent,
however, with the mechanism of the formation of H$_2$ by diffusion of
physisorbed H atoms on silicate grains, which is effective only at very low
temperatures (Pirronello et al.\ 1997). A mechanism in which H atoms are
chemically attached to the grains and in which the second H atom lands directly
on the attached H atom, the so-called Eley-Rideal mechanism (Herbst 2000)
is one possible mechanism to explain a large formation
rate at high temperatures.  Porous (carbonaceous) grains
and/or large PAH molecules appear to be the best candidate materials, but
quantitative experiments and calculations on interstellar grain analogs
at high temperatures are needed. In any case, an increase in the surface
to volume ratio over the standard grain models may be needed to allow
the high formation rates that seem to be necessary to explain our 
observations.

An alternative mechanism to increase the H$_2$($J$) populations has been
proposed by Spaans (1995), Joulain et al.\ (1998) and Falgarone et al. 
(2000)---
using the energy released by the dissipation of turbulence or by weak
(few km s$^{-1}$) shocks. The observed H$_2$ intensities toward S140
are within a factor of a few of those predicted by Joulain et al. for a cloud
with $n=30$ cm$^{-3}$. At densities typical for the periphery of 
S140, however, the turbulent model
intensities should be reduced significantly. The models of Joulain
et al.\ do not take ultraviolet pumping into account, which will enhance the
populations of the higher $J$ levels. Further development of these models and
constraints from other observed species will be needed to test them fully.

It should be noted that the strengths of the S(0) and S(1) lines observed in
the outskirts of S140 are comparable to those inferred from ultraviolet
observations of H$_2$ in diffuse and translucent clouds. Consider as an
example the recent FUSE results for the translucent cloud toward HD 73882 ($
A_{V} $=2.4 mag) by Snow et al.\ (2000). The observed column densities in $ J
=2$, 3, and 5 translate into intensities of $ 2.1\times 10^{-6} $, $ 1.4\times
10^{-5}$, and $ 1.7\times 10^{-5} $ erg s$ ^{-1} $ cm$ ^{-2}$ sr$^{-1}$ for
the S(0), S(1), and S(3) lines, respectively. The S(1) intensity is comparable
to our observed values, but that of S(0) is lower and  of S(3) higher. Indeed,
the corresponding excitation temperatures for the $ J =2$--7 levels of $ \sim
$300~K are larger than those of $ \sim 100$ K found here for the 
$J=2$--3 levels, although they are consistent with that found for the $J=3$--5
levels at one position. More sensitive searches for the higher rotational
lines of H$_2$ are needed to constrain its excitation and determine the
relative roles of ultraviolet pumping, collisions, and the H$_2$ formation
mechanism in establishing the level populations. 
FUSE studies of the H/\hh\ transition in well-characterized, low \hh\ 
column density regions are also interesting to provide further
constraints on the empirical \hh\ formation rate in different
environments using the analysis of Jura (1975).

In summary, the H$_2$ excitation temperatures and the widespread nature of
the emission suggest a PDR origin, in which the
H$_2$ lines arise from the outer warm layers. However, the absolute model
intensities fall short by an order of magnitude, unless the H$_2$ formation
rate is enhanced significantly. Other explanations invoked for high-density
PDRs, such as geometry or gas-grain drift, do not apply to these regions.
Together with the results from ultraviolet observations, our results
suggest that our understanding of the basic chemical processes 
involving H$_2$ is still incomplete.

\section{LESSONS FOR OTHER GALAXIES} \label{sec4:exgal}

The \CII\ 158 $\mu$m line and \OI\ 63 $\mu$m lines are the dominant coolants
of the interstellar gas in galaxies.  A significant fraction of the \CII\
emission from
late-type galaxies arises in dense PDRs with some additional contributions
 from atomic gas not associated with molecular clouds. In the Milky Way, the
\CII\
 emission fraction is about equally divided between  PDRs associated with
molecular
 clouds and neutral, purely atomic clouds
(Petuchowski \& Bennett
1993, Mochizuki \& Nakagawa 2000, but see Heiles 1994 for a different view).
PDRs contribute
a larger fraction of the \CII\ emission from galaxies with more active
star formation (Crawford et al. 1985,
Stacey et al. 1991) while, particularly in the outer parts of galaxies
with lower star formation rates, purely atomic gas may contribute to
the \CII\ emission (Madden et al. 1993).
The \fir\ continuum  emission from galaxies comes primarily from the same
PDRs, although, in the lowest luminosity spirals, optical emission from 
late-type stars may also contribute to the heating of dust in the molecular 
ISM (Smith et al. 1994)
and something like half the \fir\ emission may come from purely atomic clouds
(Mochizuki \& Nakagawa 2000).
It should be possible,
in principle, to derive information about the physics of the neutral ISM in
galaxies
from the \fir\ continuum and fine-structure line measurements (Wolfire,
Tielens \& Hollenbach, 1990, Stacey et al. 1991, Carral et al. 1994, 
Lord et al. 1996, Malhotra et al. 2001).

Recent studies of starburst galaxies (Luhman et al. 1998) and of a range of
lower luminosity spirals (Malhotra et al. 1997, 2001)
have made much of an apparent ``deficit'' in the \CII\ emission.
In many of the starbursts and in a fraction
of the spirals, the \CII/FIR luminosity ratio lies as much as a factor
of 10 below the canonical value (3$\times$10$^{-3}$), 
more typical for late-type galaxies.
Galactic photodissociation regions with high $n$ or high $\go$ can
have \CII/FIR ratios as low or lower than those of galaxies with the
biggest \CII\ deficits (e.g. Genzel, Harris, \& Stutzki 1989).  
The [O~I] line then provides a much larger fraction of the cooling power 
(cf. the values for the S140 cut and the S140 IRS 1 position
in Table \ref{s140lws-res}), and dust temperatures are higher. 
However, Malhotra et al. (2001) argue that the {\it total} fine structure 
emission
(\OI\ plus \CII) is decreased substantially only if $\go/n$ is increased.
This combination leads to highly charged grains and inefficient photoelectric
heating (Tielens \& Hollenbach 1985a, Bakes \& Tielens 1994).
Malhotra et al. (2001) considered but rejected a variety of explanations,
concluding that high values of $\go/n$ in the emitting regions provide
the best explanation for their sample of ``normal'' spirals.

Studies of \fir\ line emission from other galaxies have made use of 
the same models as adopted here, which
assume a uniform incident radiation field and a single PDR density for the
emitting
material (Kaufman et al. 1999,  Stacey et al. 1991, Wolfire et al. 1989).
In addition to using simple models, we can gain insight into the nature of
the regions in galaxies responsible for \fir\ fine-structure line emission by
comparing galaxy PDR emission to emission from Galactic sources with an
appropriate range of
properties.
The ISO observations of fine-structure line emission from other galaxies take
in entire giant molecular cloud complexes and the nuclear regions of galaxies.
They therefore contain contributions not only from the high excitation regions
near OB stars but also from large regions with lower UV fields.  With our
measurements
of the cut through the outer part of the S140 cloud, a comparison of Galactic
and extragalactic fine-structure line emission can include a
range of Galactic PDR conditions.
Figure 6(a-c) reproduces the plots shown in Figures 1a, 5, and 8 of 
Malhotra et al. (2001) and shows the variation of
Log$(L(\CII)/L(FIR))$ (Fig. 6a), $L(\OI)/L(\CII)$ (Fig. 6b), 
and Log$((L(\CII)+L(\OI))/L(FIR))$ (Fig. 6c) versus 
$F_{\nu}(60 \micron)/F_{\nu}(100 \micron)$
for the 45 normal, star-forming galaxies in Table 7 of Malhotra et al. (2001).
Figure 6 compares these data to the average values from our S140 cut
and to typical values for high \go\ Galactic PDR sources  (labeled \HII\
for brevity), taken from Genzel, Harris, \& Stutzki (1989) to be 
$L(\CII)/L(FIR)= 8.5\times10^{-5}$,
$L(\OI)/L(\CII)=5.8$, and $F_{\nu}(60 \micron)/F_{\nu}(100 \micron)=1.2$.
Figure 6 clearly shows that the galaxies inhabit a part of parameter
space that is different from either the S140 cut or the high \go\ PDR
sources. 

Because the galaxies generally lie between the points for the S140 cut
and the high \go\ PDRs, we attempted to reproduce the galaxy observations
by mixtures of the S140 cut and the high \go\ PDRs, indicated by the
mixing line connecting these two points.
The line represents parameters calculated by varying
the contributions of the two connected Galactic components in proportion to
the fraction of the \CII\ emission contributed by each.
It is not possible, however, to produce
the observed positions of the galaxies in the figures with any linear
combination of the S140 cut and the high \go\  PDRs. The result is similar
if we use the emission from NGC 7023, representing reflection nebulae, 
instead of the \HII\ regions.  At all values of 
$F_{\nu}(60 \micron)/F_{\nu}(100 \micron)$, 
the mixture produces too much fine-structure line emission 
(in particular, too much \CII),
when compared to the galaxy sample. At a given value of
$F_{\nu}(60 \micron)/F_{\nu}(100 \micron)$, for example, the value of
Log$((L(\CII)+L(\OI))/L(FIR))$ observed for the galaxies lies on
average 0.3 below the constructed mixture.

The results in Figure 6 show that one cannot produce the observed
emission ratios in galaxies by simply combining regions like the S140
cut with regions of active star formation in our Galaxy.  Malhotra
et al. (2001) argued that models with high $\go/n$ can explain individual
galaxies where the [CII] and [OI] is weak compared to the far-IR
continuum.  The systematically lower values of the galaxy ratios compared
to positions along our mixing line, however, suggests that the
conditions in heterogeneous clouds would not lead to appropriately weak
line fluxes at a given value of $F_{\nu}(60\micron)/F_{\nu}(100\micron)$. 
Rather than focussing solely on the line to continuum ratio, it 
may be worthwhile to examine possible reasons why the 
$F_{\nu}(60\micron)/F_{\nu}(100\micron)$ ratio might vary.
If that flux ratio increased less
rapidly along the mixing line, the data could be matched.
Mixing very small amounts of high \go\ PDR emission has a big effect on
$F_{\nu}(60 \micron)/F_{\nu}(100 \micron)$, while the effect on the line
emission is more modest, as indicated by the tick marks on the mixing line that
indicate where 90\% and 50\% of the \CII\ emission comes from the S140 cut. 
Moderating the increase in $F_{\nu}(60 \micron)/F_{\nu}(100 \micron)$ as one
mixes in small amounts of high \go\ PDR emission properties could allow the 
mixing
line to pass through most of the points.
This would require a component of emission from cooler dust.
Lowering the 60--100 $\mu$m color temperature of the high \go\ PDRs
from $\sim53$ K to $\sim35$ K (for a $\lambda ^{-1}$
emissivity law) could account for the difference. Even colder dust
is common in the molecular clouds near the center of our Galaxy
(Lis et al. 2001).

There are at least two possible causes for the cooler dust.
The dust properties in the inner few kiloparsecs of galaxies, 
where most of the ISO emission arises, might differ significantly 
from those in nearby high \go\ Galactic PDRs. 
Alternatively, there might be a contribution from regions heated by 
cooler stars or from very opaque regions in molecular clouds.
These regions would need to increase the fraction of the emission in the 
100 $\mu$m IRAS band from 25\% to 40\% of the total \fir\ luminosity.  
Whichever reason is correct, observations of some regions in the 
inner part of our own Galaxy (G0.68-0.2, ISO data archive) 
find a  \fir\ line to continuum ratio lower than for all the 
galaxies in the Malhotra sample, and yet the
$F_{\nu}(60 \micron)/F_{\nu}(100 \micron)$ is only 0.2.  
A mixing line connecting G0.68-0.2 to the \HII\ region/high \go\ PDR's
would run below the bulk of the galaxies in Figures 6a and 6c. More
complex 3-part mixtures of cold cloud edges like the S140 cut, cold
inner galaxy clouds like G0.68 and warm \HII\ regions could presumably
fill the part of parameter space occupied by the galaxies from the
Malhotra sample.
Understanding the properties
of inner Galaxy regions like G0.68 may provide further insight into the
conditions in extragalactic PDRs.

\section{SUMMARY}   \label{sec4:sum}

We have studied
the peripheral region of the molecular cloud L1204/S140, where
the FUV radiation and density are relatively low, using
ISO to observe \CII, \OI, and \hh\ lines,
and the CSO to observe the submillimeter line of atomic carbon and
millimeter emission lines of \CO\ and other CO isotopes.  
We analyzed the results with published PDR models and our own PDR models,
coupled to a Monte Carlo simulation code for radiative transfer.  
The important results from this work are:

\begin{enumerate}
\item We detected wide-spread \CII, \OI, and \hh\ emission in the region.
The \CII\ and \OI\ lines display much less spatial variation than
\CI\ and the lines from \CO\ and other CO isotopes, which reveal
the presence of a denser, shielded core that does not emit significantly 
in the \CII\ and \OI\ lines.

\item The average \CII\ and \OI\ intensities of the S140 cut are
$(8.66\pm 1.27)\times 10^{-5}$ and $(1.35\pm 0.34)\times 10^{-5}$
ergs s$^{-1}$ cm$^{-2}$ sr$^{-1}$; the average line ratio (\OI/\CII) is
$0.16\pm0.04$.
Therefore the \CII\ line is the dominant cooling line in this region.
This is quite different from what has been observed in the S140 bright PDR
by ISO, where the ratio \OI/\CII\ is $1.55\pm1.10$. 

\item Using the model predictions in Hollenbach et al.\ (1991) and the
100 \micron\ IRAS flux, the FUV intensity \go\ for the region would be 40--60,
in the situation of single-sided FUV radiation on the cloud.
If the FUV impinges on both sides, the inferred $\go \sim 20-30$.
Based on \COO\ emission and simple LVG modeling, 
the density of the gas, $n \sim 1000$ \cmv, and $\AV \sim 16$. 
Conditions in the shielded core are different: $n \sim 1\ee4$ \cmv, 
and $\AV \sim 25$.

\item The PDR models (both published and our own) indicate that the 
observed \OI\ intensity and the ratio \CII/\OI\ are best reproduced 
with $\go \sim 15$ and $n \sim 1000$ \cmv.  Our models, coupled to
a Monte Carlo simulation of the radiative transfer,
reproduce the observed \COOTWO\ emission for $\AV \sim 18$ mag.
These conditions are also consistent with the constraints on density
from line ratios of \CCO\ and \HCOP. Our models also reproduce the
observed \CI\ line, while the published models of Kaufman et al. (1999)
overproduce this line.  In this regime of $n$ and $\go$, the 
\OI\ intensity is very difficult to model because it is extremely 
subthermal and also very opaque (the optical depth is 9--14).
Trapping is very important in determining the emergent \OI\ intensity.

\item Quantitative analysis of the ionization structure and timescale for
cloud collapse for the peripheral regions of S140 
leads to conclusions consistent with a picture of 
photoionization-regulated star formation.

\item The detection of widespead  \hh\ rotational emission is not
readily explained with standard PDR models and \hh\ formation rates.
Possible solutions include a substantial enhancement in the the \hh\ formation
rates or localized heating by
weak shocks associated with turbulent decay.

\item While the emission in \CII\ and \fir\ continuum from other galaxies
lies between the characteristics of the emission from the peripheral
region studied here and regions of higher \go\ and $n$, simple mixtures
of these regions are not able to reproduce the characteristics of 
the emission from other galaxies.

\end{enumerate}

We thank the referee for a careful reading and for suggestions that
improved the paper.
We would like to extend our sincere thanks to the wonderful service by
the ISO supporting staff at IPAC (S. Lord, S. Unger, D. Levine,
L. Hermans especially) and at SRON Groningen (E. Valentijn and F. Lahuis
in particular). We also thank the CSO staff members and 
E. Gregersen and K. Mochizuki for helping
with the CSO observations.  M. Spaans and D. Jansen provided
useful discussions and help with PDR codes; E. Roueff provided data for
comparison of PDR models.
This research has made use of
NASA's Astrophysics Data System Abstract Service,
the Simbad database, operated at CDS, Strasbourg, France,
and the Online Services of IRSKY and IBIS at IPAC, and the SkyView
Online Service.
WL was partially supported by a Continuing Fellowship
and a David Bruton, Jr., Fellowship
of the University of Texas, a Frank N. Edmonds,
Jr. Memorial Fellowship and a David Alan Benfield Memorial Scholarship
of the Department of Astronomy, the University of Texas.
Work with the CSO re-imager was supported by NSF grant AST-9530695.
The research was supported by  NASA Grants NAG2-1055, NAG5-3348, and
the State of Texas, NWO grant 614.41.003, a NWO Spinoza grant, and a 
NWO bezoekersbeurs.



\clearpage

\begin{figure}
\includegraphics{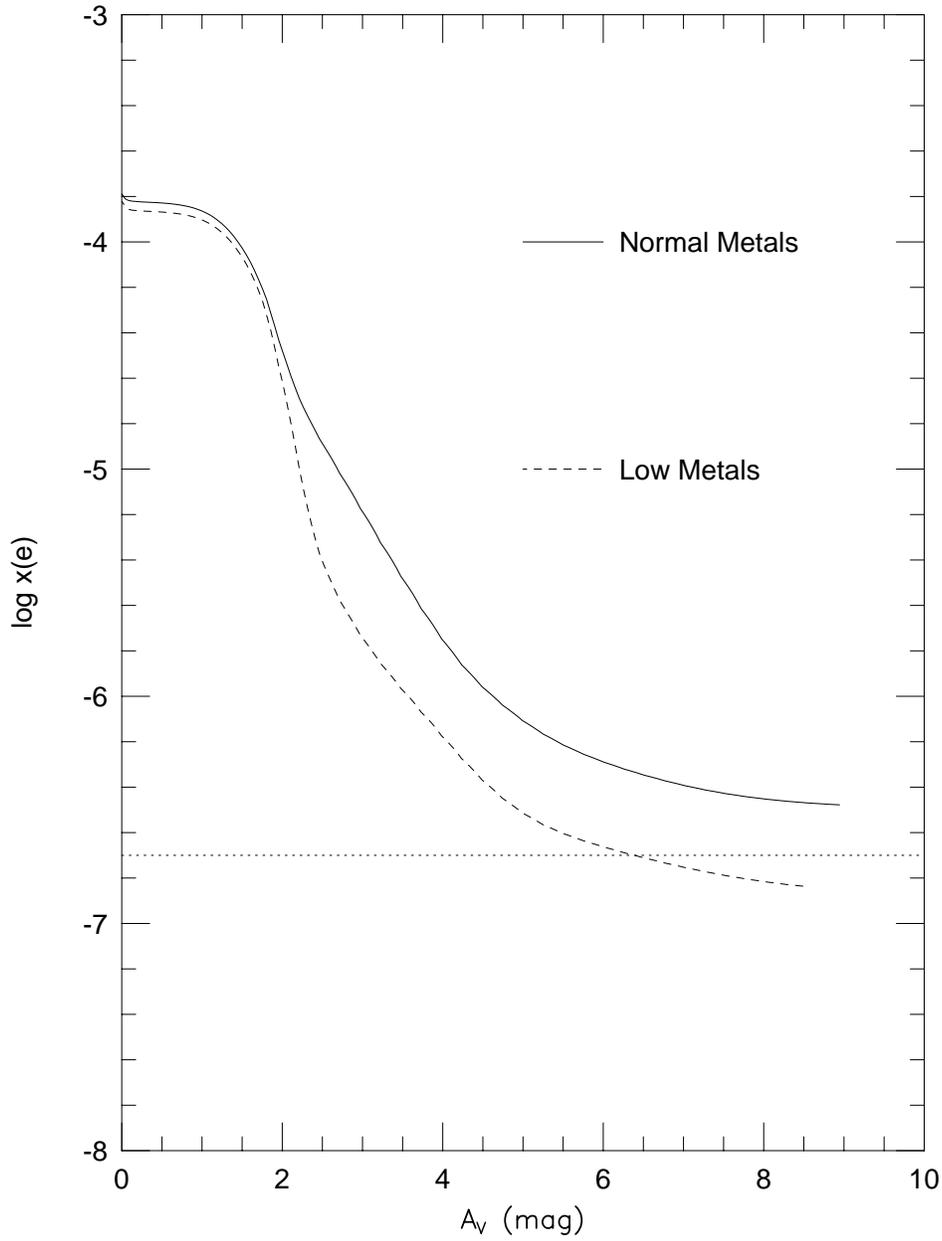}
\vskip 7.1in
\caption{Electron abundance versus extinction for two models, both with
$\go = 17$ and $n = 1000$. The solid line is for normal metals, and
the dashed line has low metals. The horizontal dotted line indicates
the level of $x(e)$ at which the ambipolar diffusion time equals
30 Myr. \label{ionize}}

\end{figure}

\begin{figure}
\includegraphics{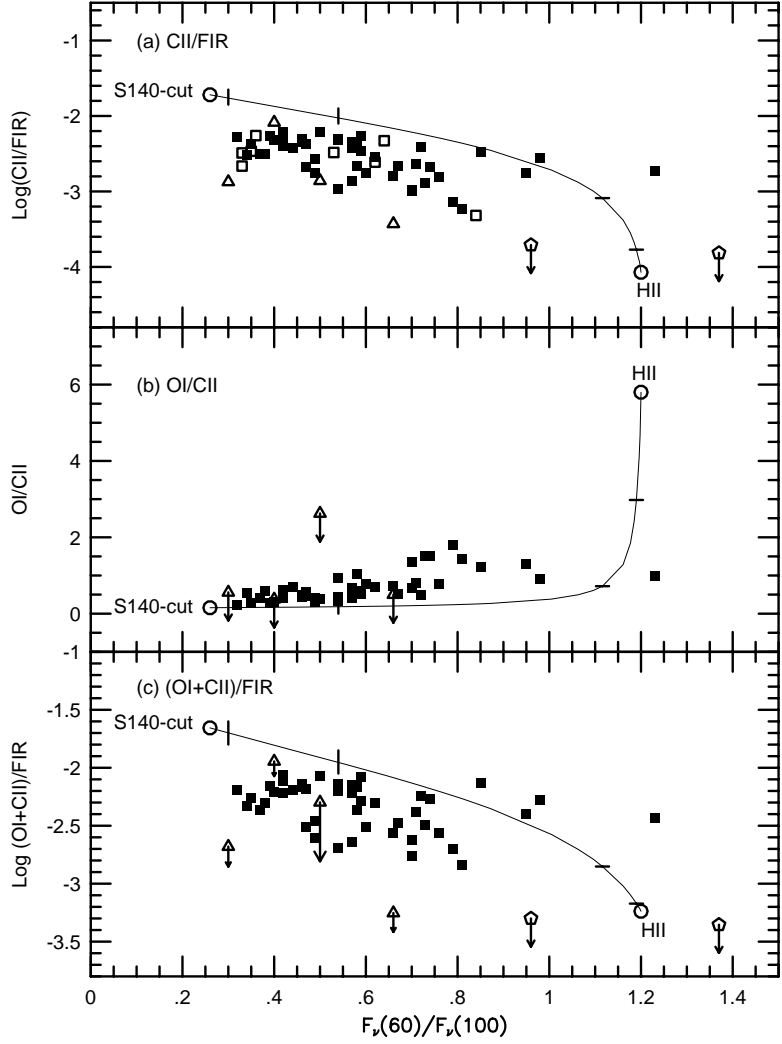}
\vskip 5.8in

\caption{
Ratios of line and continuum emission from galaxies in the study
of Malhotra et al. (2001) compared with the same ratios in Galactic sources. 
The filled squares show the ratios for galaxies
detected in both the 158 $\mu$m \CII\ line and the 63 $\mu$m \OI\ line.
Where applicable, the open squares show the ratios for galaxies observed
in \CII\ but not in \OI. The open triangles show the locations for galaxies
detected in \CII\ and observed but not detected in \OI.  In Figure 6c, the
detected \CII\ emission places a lower bound on the $(\OI + \CII)/FIR$
flux ratio. In this panel, the upper limit arrows end at the value 
corresponding to the measured $\CII/FIR$ ratio.
The open pentagons show the limits on the ratios for
galaxies observed but detected in neither \CII\ or \OI. 
In each panel, the measured ratios
for the S140 cut are shown as open circles and labeled.  The measured ratios
for the mean of a sample of high $\go$ PDRs, labeled \HII\ are also shown.  
The mixing line connecting these points is based on
the fraction of \CII\ emission contributed by regions like the S140 cut;
the horizontal tickmarks along the curves indicate where
such regions contribute 90\% and 50\% of the \CII\ emission  to the
mixture.  The vertical tickmarks show where regions like
the S140 cut contribute 90\% and 50\% of the \fir\ emission.
\label{galaxyfig}}

\end{figure}


\clearpage

\begin{deluxetable}{cc}
\footnotesize
\tablewidth{0pc}
\tablecaption{The ISO Positions (1950) of S140. \label{isopos}}
\tablehead{
\colhead{Object} & \colhead{S140} }
\startdata
Ref. Coord.  & (22\h 17\m 42\s, 63\arcdeg 23\arcmin 15$\arcsec$) \\
End Position 1 & (1368$\arcsec$, 0) \\
End Position 2 & ($-1206\arcsec$, 0) \\
Step & 99$\arcsec$ \\
CSO Off Pos. & (22\h 13\m 22\s, 63\arcdeg 23\arcmin 15$\arcsec$) \\
ISO Off Pos. & (22\h 05\m 46\fs 7, 63\arcdeg 23\arcmin 15$\arcsec$) \\
\enddata
\end{deluxetable}

\begin{deluxetable}{lrrccccc}
\footnotesize
\tablewidth{0pc}
\tablecaption{Log of CSO Observations \label{cso-obs}}
\tablehead{
\colhead{Time} & \colhead{Object} & \colhead{Transition}
               & \colhead{Frequency}
               & \colhead{$\Delta$v\tablenotemark{a}}
               & \colhead{Mode\tablenotemark{b}}
               & \colhead{$\theta_b$\tablenotemark{c}}
               & \colhead{$\eta_{mb}$\tablenotemark{d}} \\
\colhead{}     & \colhead{}       & \colhead{}
               & \colhead{(GHz)}
               & \colhead{(\kms)} & \colhead{}
               & \colhead{(arcsec)} & \colhead{} }
\startdata
June 1995 & S140 & \COTWO\ & 230.53797
                 & 0.23 & RIMG & 120 & 0.80 \\
          &      & \CCOTWO\ & 220.39870
                 & 0.24 & RIMG & 120 & 0.80 \\
          &      & \COOTWO\ & 219.56032
                 & 0.23 & RIMG & 120 & 0.80 \\
          &      & \CCOTHR\ & 330.58800
                 & 0.12 & RIMG & 110 & 0.80 \\
June 1996 & S140 & \CI\ $J=1\rightarrow0$ & 492.16070
                 & 0.08 & FD & 14 & 0.45 \\
April 1997 & S140 & \HCOPTHR\ & 267.55762
                 & 0.12 & RIMG & 160 & 0.80 \\
          &      & \HCOPTHR\ & 267.55762
                 & 0.12 & FD & 26 & 0.70 \\
\tablenotetext{a}{The velocity resolution of the data.}
\tablenotetext{b}{The setup of the observing mode.  RIMG stands for the
re-imager mode and FD the full dish.}
\tablenotetext{c}{The beam size.}
\tablenotetext{d}{The main-beam efficiency.}
\enddata
\end{deluxetable}

\begin{deluxetable}{lrrrrrcc}
\footnotesize
\tablewidth{0pc}
\tablecaption{Lines Observed by ISO \label{iso-lines}}
\tablehead{
\colhead{Line} & \colhead{Wavelength}
               & \colhead{$E_{\rm up}/k$\tablenotemark{a}}
               & \colhead{$E_{\rm low}/k$\tablenotemark{b}}
               & \colhead{$n_{\rm crit}$\tablenotemark{c}}
               & \colhead{$A$}
               & \colhead{Instr.}
               & \colhead{Beam\tablenotemark{d}}\\
\colhead{} & \colhead{(\micron)}
           & \colhead{(K)}
           & \colhead{(K)}
           & \colhead{(\cmv)}
           & \colhead{(s$^{-1}$)}
           & \colhead{} & \colhead{}
}
\startdata
\OI\ $^3$P$_1\!\rightarrow\!^3$P$_2$ & 63.183705 & 227.72 & 0 & 4.7(5)
      & 8.87(-5) & LWS & 1.3\ee{-7} \\
\CII\ $^2$P$_{3/2}\!\rightarrow\!^2$P$_{1/2}$ & 157.7409 & 91.22 & 0 & 2.8(3)
      &2.4(-6)  & LWS & 6.65\ee{-8} \\
\hh\ \jj20                         & 28.218    & 509.88 & 0  & 5.4(1)
      &2.9(-11) & SWS & 1.3\ee{-8} \\
\hh\ \jj31                         & 17.035    & 1015.12 & 170.48  & 1.1(3)
      &4.8(-10) & SWS & 8.9\ee{-9} \\
\hh\ \jj53                         & 9.662     & 2503.82 & 1015.12  & 1.9(5)
      &9.8(-9) & SWS  & 6.6\ee{-9} \\

\tablenotetext{a}{The energy of the upper state of the transition
relative the ground state in temperature.}
\tablenotetext{b}{The energy of the lower state of the transition
relative the ground state in temperature.}
\tablenotetext{c}{The critical density is $A/\gamma$, where $\gamma$ is the 
collision
rate coefficient.}
\tablenotetext{d}{The LWS beam solid angles are taken from Gry et al. (2001).}
\enddata
\end{deluxetable}

\begin{deluxetable}{crcccccccccccccc}
\footnotesize
\tablewidth{0pc}
\tablecaption{Results of CSO Observations. \label{s140cso-res}}
\tablehead{
\colhead{No.} & \colhead{Position}
               & \multicolumn{2}{c}{\CO$\,2\!\rightarrow\!1$\tablenotemark{a}}
               & \colhead{}
               & \multicolumn{2}{c}{\CCO$\,2\!\rightarrow\!1$\tablenotemark{b}}
               & \colhead{}
               & \multicolumn{2}{c}{\CCO$\,3\!\rightarrow\!2$\tablenotemark{b}}
               & \colhead{}
               & \multicolumn{2}{c}{\COO$\,2\!\rightarrow\!1$\tablenotemark{b}}
               & \colhead{}
               & \multicolumn{2}{c}{\CI\tablenotemark{b}} \\
\cline{3-4} \cline{6-7} \cline{9-10} \cline{12-13} \cline{15-16} \\
\colhead{} & \colhead{} & \multicolumn{2}{c}{(K)} & \colhead{}
           & \multicolumn{2}{c}{(K \kms)} & \colhead{}
           & \multicolumn{2}{c}{(K \kms)} & \colhead{}
           & \multicolumn{2}{c}{(K \kms)} & \colhead{}
           & \multicolumn{2}{c}{(K \kms)} }
\startdata
1 & ($-117\arcsec$, 0) &   ... &  1.1 & &  0.6 &  ... & &  ... &  ...
  & &  ... &  ... & &  ... &  ... \\
2 &  ($-18\arcsec$, 0) &   2.7 &  1.8 & &  0.8 &  ... & &  ... &  ...
  & &  ... &  ... & &  ... &  ... \\
3 &   (81$\arcsec$, 0) &   3.9 &  3.4 & &  1.2 &  ... & &  ... &  ...
  & &  0.1 &  ... & &  ... &  ... \\
4 &  (180$\arcsec$, 0) &   6.7 &  3.2 & &  3.9 &  ... & &  1.4 &  ...
  & &  0.4 &  ... & &  1.6 &  ... \\
5 &  (279$\arcsec$, 0) &   4.5 &  5.4 & &  6.2 &  ... & &  1.3 &  0.5
  & &  0.4 &  ... & &  2.8 &  ... \\
6 &  (378$\arcsec$, 0) &   4.1 &  5.8 & &  6.5 &  ... & &  0.8 &  ...
  & &  0.3 &  ... & &  3.2 &  1.5 \\
7 &  (477$\arcsec$, 0) &   7.2 &  5.1 & &  4.3 &  2.2 & &  1.1 &  0.9
  & &  0.2 &  0.1 & &  3.7 &  1.5 \\
8 &  (576$\arcsec$, 0) &   7.2 &  9.2 & &  6.4 &  5.6 & &  1.5 &  1.6
  & &  0.3 &  0.4 & &  3.7 &  4.3 \\
9 &  (675$\arcsec$, 0) &   7.7 &  9.5 & &  6.1 & 10.6 & &  1.4 &  2.6
  & &  0.5 &  1.4 & &  2.7 &  3.8 \\
10 & (774$\arcsec$, 0) &   8.6 & 11.8 & &  7.3 & 15.7 & &  2.5 &  5.7
  & &  1.2 &  2.8 & &  2.6 &  6.3 \\
11 & (873$\arcsec$, 0) &   6.5 & 11.3 & &  4.4 & 14.8 & &  2.5 &  6.2
  & &  0.4 &  2.5 & &  ... &  6.0 \\
12 & (972$\arcsec$, 0) &   6.3 & 10.2 & &  3.6 &  8.0 & &  1.7 &  0.8
  & &  0.1 &  1.4 & &  ... &  3.0 \\
13 & (1071$\arcsec$, 0) &   6.6 &  7.9 & &  3.0 &  2.4 & &  ... &  ...
  & &  ... &  0.2 & &  ... &  3.9 \\
14 & (1170$\arcsec$, 0) &   5.6 &  3.6 & &  1.3 &  1.3 & &  ... &  ...
  & &  ... &  ... & &  ... &  1.5 \\
15 & (1269$\arcsec$, 0) &   3.7 &  4.1 & &  ... &  2.7 & &  ... &  ...
  & &  ... &  ... & &  ... &  1.6 \\
16 & (1368$\arcsec$, 0) &   3.3 &  2.4 & &  ... &  1.5 & &  ... &  ...
  & &  ... &  ... & &  ... &  1.2 \\
``Cloud"\tablenotemark{c} &       & 5.6  & 4.0 &  & 4.0 & 2.0  & & 1.6  & 0.70
&  & 0.39 & 0.15 & & 2.9 & 1.9  \cr
\tablenotetext{a}{The peak T$_{\rm A}^*$ of the two velocity components
with $V_{\rm LSR}\sim -7.5$ \kms\ and $V_{\rm LSR}\sim -10.5$ \kms\
respectively.}
\tablenotetext{b}{The integrated line intensities of the two velocity
components in the region.}
\tablenotetext{c}{The mean integrated line intensities; for the $-10.5$
\kms\ component, the positions with $\delta \alpha$ between 576 and 972
were excluded.}
\enddata
\end{deluxetable}

\begin{deluxetable}{lcccc}
\footnotesize
\tablewidth{0pc}
\tablecaption{Results of the \HCOPTHR\ Line Observations.
              \label{s140-hcop}}
\tablehead{
\colhead{Position} & \colhead{FWHM Beam}
                   & \multicolumn{2}{c}{Area\tablenotemark{a}}
                   & \colhead{$\sigma$ (rms)} \\
\cline{3-4} \\
\colhead{} & \colhead{} & \multicolumn{2}{c}{(K \kms)} & \colhead{(K)}
}
\startdata
($774\arcsec$, 0) & 160 & 0.09 & 0.33 & 0.05 \\
($915\arcsec$, $-71\arcsec$)\tablenotemark{b} & 26 & ... & 5.75 & 0.41 \\

($1170\arcsec$, 0) & 160 & ... & ... & 0.04 \\
($180\arcsec$, 0) & 160 & ... & ... & 0.04 \\
\tablenotetext{a}{The integrated line intensities of the two components
are given if applicable, with the first one at $V_{\rm LSR} \sim -7.5$ \kms\
and the second at $V_{\rm LSR} \sim -10.5$ \kms.}
\tablenotetext{b}{Peak position of the
\HCOPTHR\ emission when mapped with the full dish.  The
\HCOPTHR\ emission is clumpy in the area of $300\arcsec \times 300\arcsec$
centered at ($774\arcsec$, 0).}
\enddata
\end{deluxetable}

\begin{deluxetable}{rrccccc}
\footnotesize
\tablewidth{0pc}
\tablecaption{Results of ISO LWS Observations. \label{s140lws-res}}
\tablehead{
\colhead{No.} & \colhead{Position}
              & \colhead{\CII} & \colhead{\OI} & \colhead{Ratio}
              & \colhead{IRAS(60 \micron)\tablenotemark{a}}
              & \colhead{IRAS(100 \micron)\tablenotemark{a}} \\
\cline{3-4} \cline{6-7}\\
\colhead{} & \colhead{(arcsec)} & \multicolumn{2}{c}{(10$^{-5}$ erg cm$^{-2}$
s$^{-1}$ sr$^{-1}$ )}
           & \colhead{\OI/\CII} & \multicolumn{2}{c}{(MJy sr$^{-1}$)} }
\startdata
1 & ($-$117,0) & 8.52$\pm$0.18 & 1.05$\pm$0.06 & 0.12 & 57.3 & 219 \cr
2 &  ($-$18,0) & 8.34$\pm$0.18 & 0.88$\pm$0.19 & 0.11 & 57.2  & 221 \cr
3 &   (81,0) & 8.96$\pm$0.18 & 1.84$\pm$0.12 & 0.21 & 58.8  & 220 \cr
4 &  (180,0) & 9.05$\pm$0.27 & 1.26$\pm$0.19 & 0.14 & 66.1  & 244 \cr
5 &  (279,0) & 9.05$\pm$0.27 & 1.04$\pm$0.17 & 0.12 & 61.5  & 236 \cr
6 &  (378,0) & 9.76$\pm$0.53 & 1.67$\pm$0.25 & 0.17 & 56.2  & 232 \cr
7 &  (477,0) & 7.80$\pm$0.35 & 0.81$\pm$0.18 & 0.10 & 52.9 & 226  \cr
8 &  (576,0) & 7.30$\pm$0.27 & 1.09$\pm$0.21 & 0.15 & 52.5 & 226  \cr
9 &  (675,0) & 9.58$\pm$0.27 & 1.25$\pm$0.19 & 0.13 & 54.3 & 228   \cr
10 &  (774,0) & 10.6$\pm$0.53 & 1.72$\pm$0.19 & 0.16 & 58.8 & 230  \cr
11 &  (873,0) & 8.69$\pm$0.18 & 1.81$\pm$0.23 & 0.21 & 63.4 & 227  \cr
12 &  (972,0) & 8.61$\pm$0.36 & 1.68$\pm$0.32 & 0.20 & 63.0 & 224  \cr
13 & (1071,0) & 9.58$\pm$0.44 & 1.47$\pm$0.23 & 0.15 & 58.3 & 208  \cr
14 & (1170,0) & 9.94$\pm$0.18 & 1.55$\pm$0.11 & 0.16 & 51.3 & 191  \cr
15 & (1269,0) & 7.45$\pm$0.18 & 1.44$\pm$0.19 & 0.19 & 44.8 & 164  \cr
16 & (1368,0) & 5.32$\pm$0.18 & 1.03$\pm$0.21 & 0.19 & 39.6 & 145 \cr
   & Avg(16-pos)\tablenotemark{b} & 8.66$\pm$1.27 & 1.35$\pm$0.34
   & 0.16$\pm$0.04   & &  \cr
   & ISO-OFF    & 1.69$\pm$0.18 & 0.46$\pm$0.20 & 0.27 & 9.6 &  47.7 \cr
   & IRS1\tablenotemark{c}       & 44$\pm$2 & 68$\pm$3  & $1.55\pm0.10$ & &    
    \cr

\enddata
\tablenotetext{a}{The IRAS fluxes at 60 \micron\ and 100 \micron\ at
the S140 cut positions and the S140 ISO off position.}
\tablenotetext{b}{The average line fluxes and ratios of all 16 positions. The
errors are the standard deviation of the values at the 16 positions.}
\tablenotetext{c}{The average of the 4 existing measurements toward this 
position. The error reflects the standard deviation of the individual 
measurements.}
\end{deluxetable}

\begin{deluxetable}{rrccccc}
\footnotesize
\tablewidth{0pc}
\tablecaption{Results of ISO SWS Observations. \label{s140sws-res}}

\tablehead{
              \colhead{No.}
              & \colhead{Position}
              & \colhead{\hh\ \jj20}
              & \colhead{\hh\ \jj31}
              & \colhead{\hh\ \jj53}
              & \colhead{$T_{32}$\tablenotemark{a}} 
              & \colhead{$T_{53}$\tablenotemark{a}} 
\\
\cline{3-5} \\
\colhead{} & \colhead{(arcsec)} &
\multicolumn{3}{c}{(10$^{-5}$ erg cm$^{-2}$ s$^{-1}$ sr$^{-1}$ )}
&\colhead{(K)}
&\colhead{(K)}
}
\startdata
1 & ($-117$,0) & 1.0$\pm$0.5 & 1.1$\pm$0.6 & $<0.6$ &  $109\pm30$ & $<320$ \cr
``F'' & (743,0) & 0.8$\pm$0.4 & 0.8$\pm$0.3 & $<0.6$ & $106\pm25$ & $<320$ \cr
14    & (1170,0) & 1.3$\pm$0.4 & 0.9$\pm$0.4 & $0.8\pm0.3$ & $98\pm18$ & 
$350\pm50$ \cr

\enddata
\tablenotetext{a}{The excitation temperature between upper $J$ levels
that yields the observed
line ratio, assuming that the ortho-para ratio is in LTE.}
\end{deluxetable}


\end{document}